\documentclass[a4paper,11pt]{article}
\pdfoutput=1 

\usepackage{jheppub} 
\usepackage{subcaption}
\DeclareMathOperator{\sech}{sech}
\usepackage{tikz}
\usetikzlibrary{decorations.pathmorphing}
\usetikzlibrary{shadows.blur}
\usetikzlibrary{intersections}
\usepackage[T1]{fontenc} 
\usepackage{braket}

\def\be{\begin{equation}}
\def\ee{\end{equation}}

\usepackage{xcolor}

\renewcommand{\d}{\mathrm{d}}

\newcommand{\BUket}[6]{%
 \begin{scope}[shift={(#1,#2)}, yscale={\ifnum#3=1 -#4\else #4\fi}, xscale=#4]
 \ifnum#6=1
 \draw[dotted, thick] (0,0) ellipse (.5 and .25);
 \draw[dotted, thick] (3,0) ellipse (.5 and .25);
 \draw[thick] (0.5,0) arc[start angle=0, end angle=180, radius=0.5];
 \draw[thick] (3.5,0) arc[start angle=0, end angle=180, radius=0.5];
 \else
 \draw[thick] (0,0) ellipse (.5 and .25);
 \draw[thick] (3,0) ellipse (.5 and .25);
 \fi

 \ifcase#5
 \draw[thick,dashed] (1.5,0) ellipse (.3 and .15); 
 \or
 \draw[double,thick] (1.5,0) ellipse (.3 and .15); 
 \or
 \draw[very thick,green] (1.5,0) ellipse (.3 and .15); 
 \or
 \draw[thick] (1.5,0) ellipse (.3 and .15); 
 \fi

 \draw[thick] (3.5,0) arc[start angle=0, end angle=-180, radius=2];
 \draw[thick] (2.5,0) arc[start angle=0, end angle=-180, radius=1];
 \draw[thick, dashed] (1.2, 0) to[out = -90, in =80] (1,-1.5);
 \draw[thick, dashed] (1.8, 0) to[out = -90, in =100] (2,-1.5);
 \end{scope}%
}

\definecolor{inferencepurple}{RGB}{112,48,160}

\title{Relic of the Kink Shape Mode in a Topological Thick Brane - some implications}

\author[]{Gahan Chattopadhyay and}
\author[]{Soumitra Sengupta}

\affiliation[]{School of Physical Sciences, Indian Association for the Cultivation of Science, India}

\emailAdd{gahanch080@gmail.com}
\emailAdd{tpssg@iacs.res.in}

\abstract{The kink soliton provides a simple construction of a topological thick 3-brane sourced by a minimally coupled canonical scalar field in five-dimensional general relativity. Although the flat-space kink is a solution of $\varphi^4$ theory, the gravitating configuration requires $M_*$-suppressed $\varphi^6$ corrections to support a stable warped geometry. Unlike its flat-space counterpart, the scalar perturbation spectrum contains no normalizable bound state. We nevertheless identify a single quasi-bound, or resonant, mode. Using a time-independent relative-probability analysis, direct time evolution, and a four-turning-point WKB calculation, we establish a direct connection between this resonance and the massive bound shape mode of the flat-space $\varphi^4$ kink. We also show that, within an EFT-consistent parameter window, the resonance can transfer energy from bulk geometric fluctuations to localized matter, thereby allowing the possibility of high-scale reheating in the early universe.}

\begin{document}

\maketitle
\flushbottom
\def\A{{\mathcal A}}
\def\var{{\mathrm {var}}}
\def\OO{{\mathbb{O}}}

\section{Introduction}
\label{sec:intro}

The possibility that spacetime contains dimensions beyond the observed three spatial directions has accompanied attempts to unify the fundamental interactions for more than a century. Kaluza showed that five-dimensional general relativity contains four-dimensional gravity together with a vector field, while Klein supplied the compactification and quantum interpretation of the additional coordinate \cite{Kaluza:1921tu,Klein:1926tv}. Extra dimensions later became an intrinsic ingredient of string theory and M-theory, and they remain useful at the phenomenological level because geometry can reorganize four-dimensional mass scales, couplings, and symmetry-breaking patterns. Modern extra-dimensional models therefore address questions ranging from the hierarchy between the electroweak and Planck scales to the localization of matter and the modification of gravity at short or cosmological distances; see, for example, the reviews \cite{Maartens:2010ar,Dzhunushaliev:2009va}.

The braneworld paradigm gives a particularly economical realization of this idea: observable matter is concentrated on a $(3+1)$-dimensional hypersurface embedded in a higher-dimensional bulk, whereas gravity probes the full spacetime. The domain-wall proposal of Rubakov and Shaposhnikov provided an early field-theoretic mechanism for confining low-energy physics to a defect \cite{Rubakov:1983bb}. The large-extra-dimension scenario of Arkani-Hamed, Dimopoulos, and Dvali lowered the fundamental gravitational scale by diluting gravity through a large compact volume \cite{Arkani-Hamed:1998jmv,Antoniadis:1998ig}. Randall and Sundrum subsequently showed that an exponentially warped fifth dimension can generate the electroweak-Planck hierarchy and can localize four-dimensional gravity even when the extra dimension is noncompact \cite{Randall:1999ee,Randall:1999vf}. These constructions transformed extra dimensions from a unification device into a framework with concrete implications for particle physics, gravity, and cosmology \cite{Megias:2026wwp, Karmakar:2026bdm, Karmakar:2026rij, Waseem:2025kuu, Afrasiar:2023nir, Sengupta:2023sua, Moreira:2026tea, Dudas:2025ubq}.

The original Randall-Sundrum models describe the brane as an infinitely thin source. This idealization is often sufficient at distances much larger than the brane width, but the delta-function stress tensor does not explain the microscopic origin or internal dynamics of the brane. Thick-brane models replace it by a smooth gravitating defect, most commonly a domain wall generated by one or more bulk scalar fields. Such backgrounds can be constructed through first-order or superpotential methods \cite{DeWolfe:1999cp,Gremm:1999pj}, can reproduce localized four-dimensional gravity, and provide a field-theoretic setting in which the localization of fermions and other bulk fields may be studied \cite{Kehagias:2000au,Ringeval:2001cq,Liu:2009ve}. The thick-brane literature now includes multi-field walls, defects with internal structure, noncanonical scalar sectors, and models based on modified gravity \cite{Bazeia:2005hu,Dzhunushaliev:2009va, Tan:2024qij, Tan:2024url, Tan:2023cra}. These generalizations are valuable for engineering spectra and localization properties, but they can also obscure which phenomena are generic consequences of gravitationally dressing a simple topological defect.

A canonical kink generated by a single scalar field is therefore an especially useful benchmark. The scalar interpolates between disconnected vacua as the extra-dimensional coordinate crosses the wall, giving the brane a smooth stress-energy profile and a topological origin. For the kink profile considered here, the superpotential construction determines both the warp factor and the scalar potential analytically \cite{Kehagias:2000au}. Gravity modifies the flat-space double-well potential by a term suppressed by the five-dimensional Planck scale; in the decoupling limit this correction disappears and the ordinary $\varphi^4$ theory is recovered. This construction has several advantages over more elaborate thick branes: it uses only Einstein gravity and one canonical scalar, its background is analytic, its topology is transparent, and its flat-space limit is controlled. Most importantly for the present work, the same limit provides an unambiguous reference spectrum against which the excitations of the gravitating brane can be compared. Stability of such scalar-generated thick branes and the localization properties of their perturbations have been studied extensively \cite{Giovannini:2001fh,Kobayashi:2001jd,Kakushadze:2000zp}.

The flat-space $\varphi^4$ kink is itself one of the basic nonperturbative solutions of relativistic field theory \cite{Rajaraman:1982is,Manton:2004tk}. Its topological charge prevents continuous relaxation to the vacuum, while its fluctuation operator is an exactly solvable modified P\"oschl-Teller problem. The discrete spectrum contains the translational zero mode and one massive internal excitation, conventionally called the shape mode. The latter describes a localized oscillation of the kink profile and supplies an internal channel in which energy can be stored. Excitation of this mode underlies the standard resonant-energy-exchange explanation of the multi-bounce escape windows in $\varphi^4$ kink-antikink collisions \cite{Sugiyama:1979mi,Campbell:1983xu}. Beyond linear order, the wobbling kink radiates through its coupling to continuum modes \cite{Manton:1996ex}, and kink-radiation interactions can display effects such as negative radiation pressure \cite{Forgacs:2008az}. The shape mode is thus not merely a spectral curiosity: it is central to the nonlinear dynamics of kink collisions. In Section \ref{sec:a}, therefore, we briefly review the flat-space kink solution and its linearized fluctuation spectrum.

Once the kink gravitates, however, the scalar fluctuation cannot be varied independently of the geometry. Scalar-field and scalar-metric perturbations mix, gauge freedom must be handled explicitly, and the physical degree of freedom is described by a constrained master equation \cite{Giovannini:2001fh,Kobayashi:2001jd}. In regular warped backgrounds with finite four-dimensional Planck mass, the scalar zero mode is generally not localized, while the massive sector forms a continuum. Earlier work established that discrete modes of additional matter fields in the gravity-free domain-wall problem may become resonances after warped gravity is included \cite{DaviesGeorge:2007domainwall}; resonant Kaluza-Klein modes of bulk fields and their time evolution have also been studied in a variety of thick-brane backgrounds \cite{Liu:2009ve,Tan:2022uex}. These results demonstrate the general importance of quasi-localization, but they do not by themselves establish its connection to the internal modes of the non-gravitating scalar configuration. During the preparation of this manuscript, a study of gravitating kink-antikink collisions found independent evidence that, in two-dimensional dilaton gravity, a self-gravitating $\varphi^4$ shape mode becomes a long-lived quasi-bound state at weak gravitational coupling \cite{He:2026mvn}. Our work instead addresses how the bound shape mode becomes a quasi-localized excitation when the kink gravitates in $(4+1)$ dimensions and examines its consequences for the thick-brane geometry sourced by the kink. In Section \ref{sec:c}, we formulate the scalar perturbations of the coupled Einstein-scalar system, locate the resonance using the relative-probability method, and extract its mass and spectral width. In Section~\ref{sec:d}, we establish its shape-mode origin by showing that its mass and reconstructed scalar profile approach their flat-space counterparts as the gravitational coupling is reduced. In Section \ref{sec:f}, we evolve the resonant profile directly in time, using absorbing layers and controlled numerical dissipation to follow its oscillation and leakage into the bulk. Finally, in Section \ref{sec:add1}, a four-turning-point WKB analysis relates the width to the barrier action and explains, using a simplified model, why the resonance becomes exponentially long-lived in the flat-space limit. To our knowledge, this combined spectral, profile, time-domain, and WKB identification of the brane-forming kink's own shape mode has not previously been demonstrated in the minimal five-dimensional Einstein-scalar system.

The resonance also links bulk geometry to four-dimensional phenomenology. In Section~\ref{sec:implications}, we study the evolution of its localized energy and examine whether it can decay into matter confined within the brane thickness. Because the mode contains both scalar and metric perturbations, it couples to the trace of the localized matter stress tensor and can transfer energy from the geometric excitation to massive brane fields. The competition between this channel and leakage into the noncompact bulk depends on both the tunneling enhancement and the cutoff-sensitive factor $(k/M_*)^5$. We therefore do not assume that reheating is automatically efficient. Instead, we identify an EFT-consistent parameter window in which decay into localized matter dominates while the wall scale, resonance mass, and asymptotic AdS curvature remain below the five-dimensional cutoff. Within this window, the mechanism can support high-scale reheating and may be compatible with thermal leptogenesis \cite{Davidson:2002qv,Buchmuller:2005eh}, subject to the production abundance of the resonance and efficient thermalization of its decay products.

\section{The \texorpdfstring{$\varphi^4$}{phi4} kink soliton and its excitations}
\label{sec:a}

\begin{figure}[htbp]
    \centering

    \begin{subfigure}{0.45\linewidth}
        \centering
        \includegraphics[width=\linewidth, height=6cm]{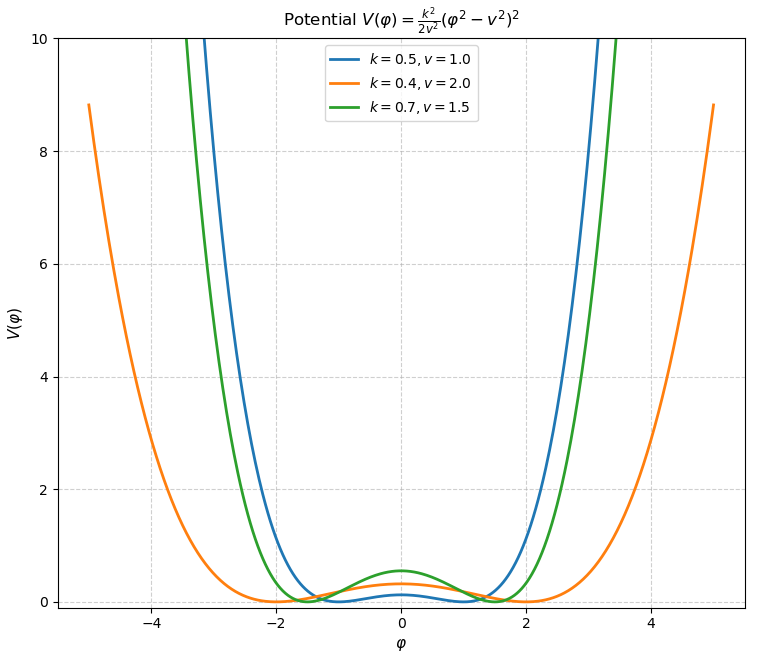}
        \caption{}\label{fig:phi4-potential}
    \end{subfigure}
    \hspace{0.1cm}
    \begin{subfigure}{0.45\linewidth}
        \centering
        \includegraphics[width=\linewidth, height=6cm]{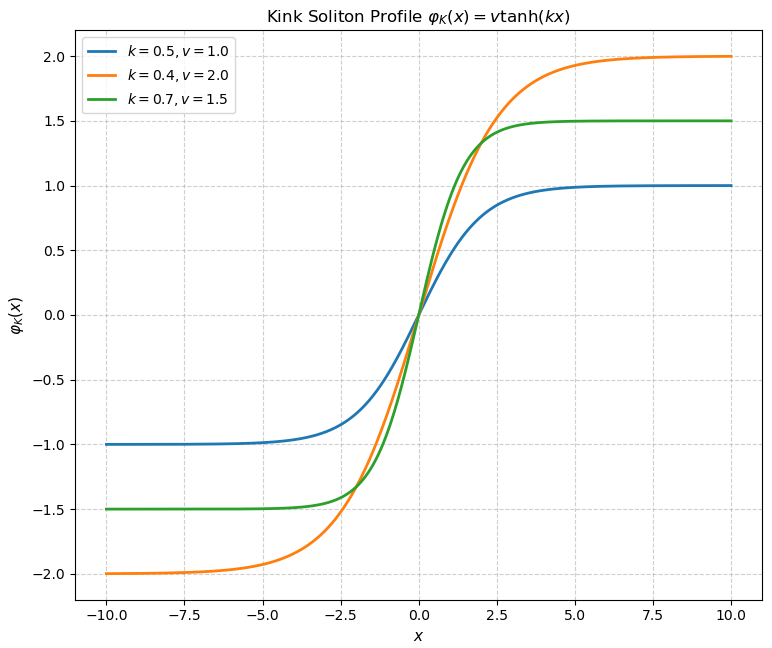}
        \caption{}\label{fig:phi4-kink}
    \end{subfigure}
    \caption{(a) Double-well potential of the $\varphi^4$ theory. (b) Corresponding kink profiles.}
    \label{fig:PotAndKink}
\end{figure}

In this section, we briefly review some basic concepts related to the $\varphi^4$-theory in flat space and its linear fluctuation spectrum. Consider a real scalar field $\varphi$ in $(1+1)$-dimensional Minkowski spacetime, with metric signature $(-,+)$ and Lagrangian density
\be\label{eq:phi4} \mathcal{L}=-\frac{1}{2}\eta^{\mu\nu}\partial_\mu \varphi\partial_\nu \varphi - V(\varphi), \ee
where
\be\label{eq:phi4pot} V(\varphi)=\frac{k^2}{2v^2}\left(\varphi^2-v^2\right)^2, \ee
and $k,v>0$. The theory is invariant under $\varphi\to-\varphi$, and the double-well potential has two degenerate vacua at $\varphi=\pm v$, as shown in Figure~\ref{fig:PotAndKink}(\subref{fig:phi4-potential}). It admits a static, finite-energy kink solution that interpolates between these vacua:
\be\label{eq:kink} \varphi_K(x)=v\tanh{\left(kx\right)}. \ee
The kink profile is shown in Figure~\ref{fig:PotAndKink}(\subref{fig:phi4-kink}).

\noindent To study the kink's excitations, we consider small fluctuations $\eta(x,t)$ about the static background:
\be\label{eq:fluc} \varphi(x,t)=\varphi_K(x)+\eta(x,t). \ee
Keeping only terms linear in $\eta$ gives the linearized Klein--Gordon equation
\be\label{eq:kg} \left[-\partial_t^2+\partial_x^2 - V''(\varphi_K)\right]\eta(x,t)=0, \ee
where $V''$ denotes the second derivative of the potential with respect to $\varphi$. Using the normal-mode ansatz $\eta(x,t)=e^{i\omega t}f(x)$, we obtain the Schr\"{o}dinger eigenvalue equation
\be\label{eq:kgSch} \left[-\frac{d^2}{dx^2}+U(x) \right]f(x)=\omega^2 f(x), \ee
with effective potential
\be\label{eq:ux} U(x)=V''(\varphi_K(x))=-2k^2+6k^2\tanh^2(kx). \ee

\noindent Introducing the dimensionless variables
\begin{equation}
    \bar x=kx, \qquad \bar\omega=\frac{\omega}{k}, \qquad
    \bar U(\bar x)=\frac{U(\bar x/k)}{k^2}=6\tanh^2\bar x-2,
\end{equation}
we can write the eigenvalue equation as
\begin{equation}\label{eq:pert}
    \left[-\frac{d^2}{d\bar x^2}+6\tanh^2\bar x-2\right]f(\bar x)
    =\bar\omega^2 f(\bar x),
\end{equation}
where, for simplicity, we retain the symbol $f$ for the mode function expressed in terms of $\bar x$.

\begin{figure}[htbp]
    \centering

    \begin{subfigure}{0.45\linewidth}
        \centering
        \includegraphics[width=\linewidth, height=6cm]{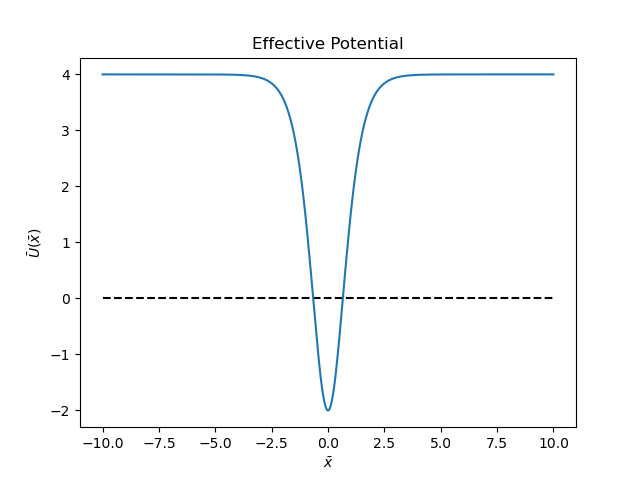}
        \caption{}\label{fig:phi4-effective-potential}
    \end{subfigure}
    \hspace{0.1cm}
    \begin{subfigure}{0.45\linewidth}
        \centering
        \includegraphics[width=\linewidth, height=6cm]{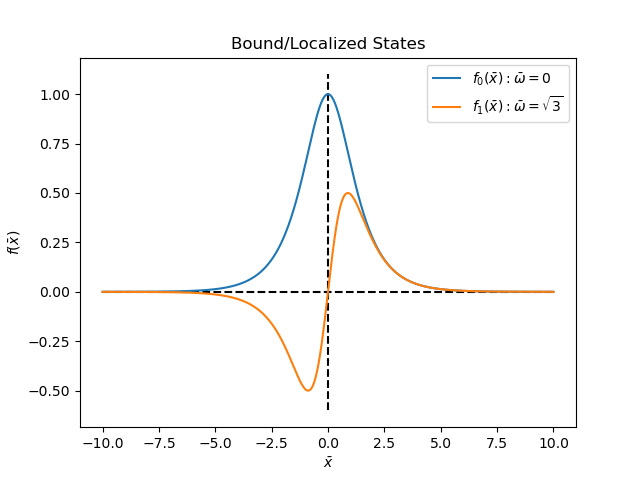}
        \caption{}\label{fig:phi4-bound-modes}
    \end{subfigure}
    \caption{(a) Effective potential for linear fluctuations about the kink. (b) The zero mode and shape mode, which form the discrete bound-state spectrum.}
    \label{fig:phi4-spectrum}
\end{figure}

\noindent The effective potential is of the modified P\"{o}schl--Teller form and admits exactly two normalizable bound states \cite{Rajaraman:1982is,Manton:2004tk}. Their frequencies and unnormalized mode functions are
\begin{align}
    \bar\omega_0&=0, & f_0(\bar x)&=\sech^2\bar x, \\
    \bar\omega_1&=\sqrt{3}, & f_1(\bar x)&=\sech\bar x\tanh\bar x.
\end{align}
The zero mode $f_0$ is even in $\bar x$ and corresponds to an infinitesimal translation of the kink. The odd mode $f_1$, known as the \textit{shape mode}, describes a localized oscillation of the kink profile. In addition to these discrete modes, the spectrum contains a continuum with $\bar\omega^2\geq4$. The effective potential and the two bound-state profiles are shown in Figure~\ref{fig:phi4-spectrum}.

\noindent The two bound modes play distinct roles in kink dynamics. The zero mode reflects continuous spatial translation symmetry: shifting the kink's position leaves its static energy unchanged and produces no restoring force. The shape mode, by contrast, is an internal vibrational degree of freedom that can store energy in localized profile oscillations. During kink--antikink collisions, energy exchange between translational motion and this internal mode provides a mechanism for resonance escape windows, in which the pair separates after multiple impacts \cite{Rajaraman:1982is, Sugiyama:1979mi, He:2026mvn, Campbell:1983xu}.

Beyond the linear approximation in equation~\eqref{eq:kg}, nonlinear coupling to continuum modes leads to radiative damping of the shape-mode oscillation \cite{Manton:1996ex}. Related nonlinear radiation effects include negative radiation pressure \cite{Forgacs:2008az}. The localized shape mode of the flat-space kink therefore provides the starting point for our investigation of its fate when the kink generates a gravitating thick brane.

\section{A thick brane generated by the kink}\label{sec:b}
A scalar kink can generate a thick 3-brane in $(4+1)$-dimensional spacetime. Here we review the construction of a gravitating kink following \cite{Kehagias:2000au} and establish the conventions used throughout the remainder of this article. Related thick-brane constructions are discussed in \cite{DeWolfe:1999cp,Gremm:1999pj,Bazeia:2005hu,Dzhunushaliev:2009va}.

We consider five-dimensional Einstein gravity coupled to a real scalar field $\varphi$, with action
\be\label{eq:action} S=\int d^5x \sqrt{-G}\left(2 M_*^3 R - \frac{1}{2}\partial_M \varphi \partial^M \varphi - V(\varphi)\right), \ee
where $M_*$ is the five-dimensional Planck scale and $G$ is the determinant of the metric $G_{MN}$. Capital Latin indices $M,N$ run over $0,1,2,3,4$, while Greek indices $\mu,\nu$ run over the four brane coordinates $0,1,2,3$. The scalar field sources the thick brane; its microscopic origin is not specified here.

We adopt a warped metric with four-dimensional Poincar\'e symmetry,
\be\label{eq:metricBg} ds^2=G_{MN}dx^M dx^N = e^{2A(y)}\eta_{\mu\nu}dx^\mu dx^\nu + dy^2, \ee
where $\eta_{\mu\nu}$ has signature $(-,+,+,+)$ and the extra dimension is noncompact, $y\in\mathbb{R}$. The function $A(y)$ determines the warp factor $e^{2A(y)}$. We take the background scalar field to depend only on $y$.

Varying the action with respect to the metric and scalar field gives the background equations
\begin{align}
    \frac{1}{2}(\varphi')^2 - V(\varphi)&=24M_*^3(A')^2, \label{eq:eoms}\\
    \frac{1}{2}(\varphi')^2 + V(\varphi)&=-12M_*^3A''-24M_*^3(A')^2, \label{eq:eoms2}\\
    e^{-2A}\Box\varphi+\varphi''+4A'\varphi'&=\frac{\partial V}{\partial\varphi}. \label{eq:eoms3}
\end{align}
Here a prime denotes differentiation with respect to $y$, and $\Box\equiv\eta^{\mu\nu}\partial_\mu\partial_\nu$ is the four-dimensional Minkowski d'Alembertian. For the background ansatz, $\Box\varphi=0$. Only two of these three background equations are independent.

We choose the scalar profile to have the kink form introduced in Section~\ref{sec:a}:
\be\label{eq:scalarBg} \varphi(x^\mu,y)=\varphi_0(y)=v\tanh(ky). \ee
In five dimensions, a canonically normalized scalar field has mass dimension $3/2$; hence $v$ has mass dimension $3/2$, while $k$ has mass dimension $1$. Rather than specifying $V(\varphi)$ in advance, we reconstruct a potential that supports this profile using the \textit{superpotential method}. This method is inspired by supersymmetry but does not require the theory itself to be supersymmetric.

Introducing an auxiliary function $W(\varphi)$, with $W_\varphi\equiv dW/d\varphi$, we impose
\begin{equation}\label{eq:super}
    \varphi'(y)=W_\varphi, \qquad
    V(\varphi)=\frac{1}{2}W_\varphi^2-\frac{1}{6M_*^3}W^2.
\end{equation}
The Einstein equations~\eqref{eq:eoms} and~\eqref{eq:eoms2} are then satisfied by the first-order equation
\be\label{eq:warp} A'(y)=-\frac{1}{12M_*^3}W. \ee
For the kink profile in equation~\eqref{eq:scalarBg}, choosing the integration constants so that $W(0)=0$ and $A(0)=0$ yields
\begin{align}
    V(\varphi)&=\frac{k^2}{2v^2}(\varphi^2-v^2)^2
    -\frac{k^2}{54v^2M_*^3}\varphi^2(\varphi^2-3v^2)^2, \label{eq:vsol}\\
    A(y)&=-\frac{v^2}{36M_*^3}\log\!\left[\cosh^2(ky)\right]
    -\frac{v^2}{72M_*^3}\tanh^2(ky). \label{eq:Asol}
\end{align}

In the flat-space limit, $M_*\to\infty$ with $v$ and $k$ held fixed, the gravitational correction to $V(\varphi)$ vanishes and the $\varphi^4$ double-well potential is recovered \cite{Kehagias:2000au}. At fixed $y$, $A(y)\to0$, so the metric becomes Minkowski. This limit connects the gravitating configuration to the flat-space kink discussed in Section~\ref{sec:a} and provides a reference point for the fluctuation analysis.

\begin{figure}[htbp]
    \centering

    \begin{subfigure}{0.3\linewidth}
        \centering
        \includegraphics[width=\linewidth, height=4cm]{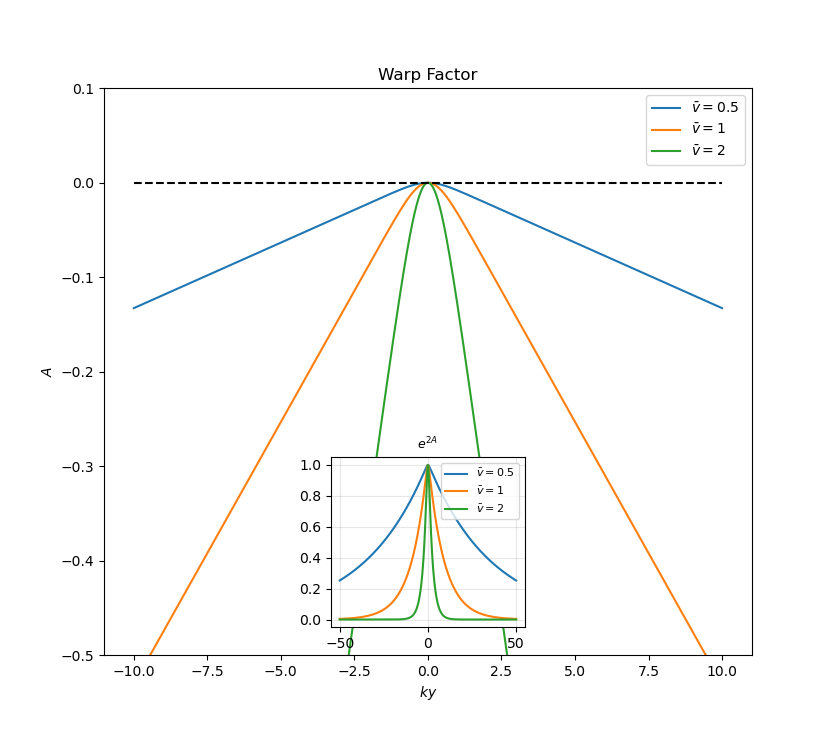}
        \caption{}\label{fig:kink-brane-warp}
    \end{subfigure}
    \hspace{0.1cm}
    \begin{subfigure}{0.3\linewidth}
        \centering
        \includegraphics[width=\linewidth, height=4cm]{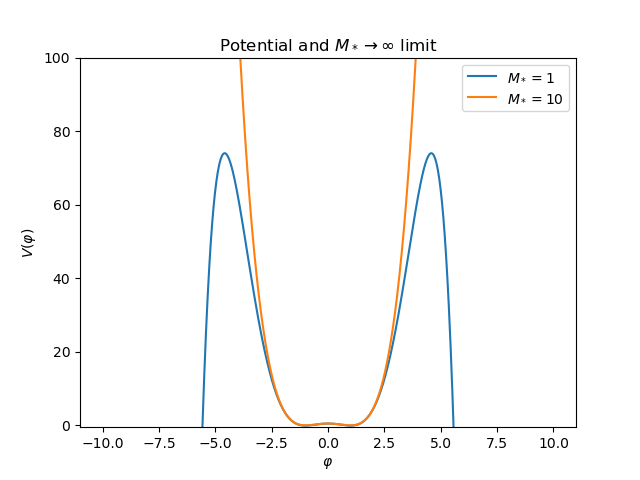}
        \caption{}\label{fig:kink-brane-potential}
    \end{subfigure}
    \hspace{0.1cm}
    \begin{subfigure}{0.3\linewidth}
        \centering
        \includegraphics[width=\linewidth, height=4cm]{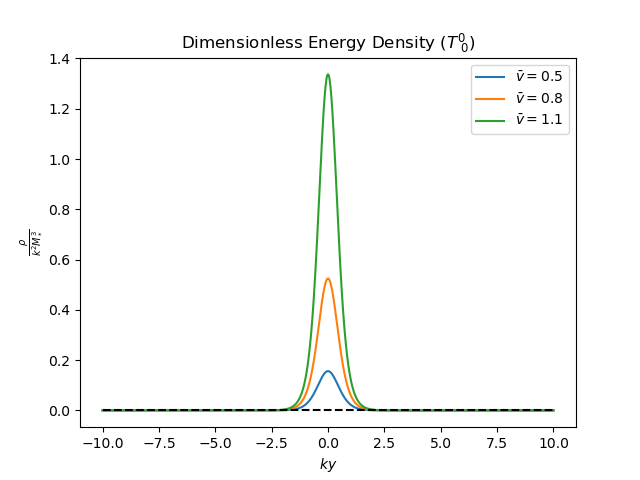}
        \caption{}\label{fig:kink-brane-density}
    \end{subfigure}
    \caption{(a) Warp profile and (c) energy density of the thick brane for different values of $\bar v=v/M_*^{3/2}$. (b) Scalar potential at large $M_*$, illustrating its approach to the $\varphi^4$ double-well form in the flat-space limit.}
    \label{fig:kink-brane-background}
\end{figure}

Integrating over the extra dimension gives the effective four-dimensional Planck scale. With the four-dimensional Einstein--Hilbert term normalized as $2M_{\mathrm{Pl}}^2R_4$, the relation is
\be\label{eq:Mpl} M_{\mathrm{Pl}}^2=M_*^3\int_{-\infty}^{\infty}dy\,e^{2A(y)}. \ee
Figure~\ref{fig:kink-brane-background} illustrates the warp profile, the approach of the scalar potential to its flat-space form, and the brane energy density.

\newpage
\section{Scalar perturbations of the thick brane}\label{sec:c}
We now extend the fluctuation analysis of Section~\ref{sec:a} to the gravitating kink. Unlike in the flat-space problem, scalar-field fluctuations couple to the scalar sector of the metric perturbations. Since these perturbations depend on the coordinate choice, the analysis requires either gauge fixing or a gauge-invariant formulation. We follow \cite{Kobayashi:2001jd} and work in the longitudinal gauge.

Define the conformal coordinate by
\be\label{conformal} z=\int_0^y e^{-A(\tilde y)}\,d\tilde y, \ee
with the brane center at $z=0$. Since $dz/dy=e^{-A(y)}>0$, the transformation is monotonic and can be inverted numerically. The background metric becomes
\be\label{eq:metricZ} ds^2=G_{MN}dx^M dx^N=e^{2A(y(z))}\left(\eta_{\mu\nu}dx^\mu dx^\nu+dz^2\right). \ee
Throughout the perturbation analysis, primes denote derivatives with respect to $z$, and we define
\begin{equation}
    \mathcal{H}\equiv e^{-A}\frac{d}{dz}e^A=\frac{dA}{dz},
    \qquad \Box\equiv\eta^{\mu\nu}\partial_\mu\partial_\nu.
\end{equation}
The derivatives $\nabla_\mu$ below refer to the flat brane metric and coincide with $\partial_\mu$ in Cartesian coordinates.

The general linear perturbation restricted to the scalar sector is
\begin{align}
    ds^2&=e^{2A(y(z))}\Bigl[\bigl\{(1+2\psi)\eta_{\mu\nu}
    -\nabla_\mu\nabla_\nu E\bigr\}dx^\mu dx^\nu \nonumber\\
    &\hspace{3.5cm}-2\nabla_\mu B\,dz\,dx^\mu+(1+2\phi)dz^2\Bigr],\\
    \varphi&=\varphi_0+\delta\varphi,
\end{align}
where $\varphi_0(z)=v\tanh(ky(z))$ is the background profile. The metric perturbations $\psi$, $\phi$, $E$, and $B$, as well as $\delta\varphi$, depend on $x^\mu$ and $z$. In the longitudinal gauge, $E=B=0$, so
\be\label{longt} ds^2=e^{2A(y(z))}\left[(1+2\psi)\eta_{\mu\nu}dx^\mu dx^\nu+(1+2\phi)dz^2\right]. \ee

For the background and linearized equations below, we use units in which $M_*=1$. The background equations in conformal coordinates are
\begin{align}
    \frac{1}{2}(\varphi_0')^2-e^{2A}V(\varphi_0)&=24\mathcal{H}^2, \label{eq:eomsZ}\\
    \frac{1}{2}(\varphi_0')^2+e^{2A}V(\varphi_0)&=-12\mathcal{H}'-12\mathcal{H}^2, \label{eq:eomsZ2}\\
    \varphi_0''+3\mathcal{H}\varphi_0'&=e^{2A}\frac{\partial V}{\partial\varphi_0}, \label{eq:eomsZ3}
\end{align}
where $\Box\varphi_0=0$. To first order in the perturbations, the Einstein and scalar-field equations give
\begin{align}
    (z,z):\quad &3\Box\psi+12\mathcal{H}\psi'-12\mathcal{H}^2\phi
    =\frac{1}{4}\left(\varphi_0'\delta\varphi'-\phi(\varphi_0')^2
    -e^{2A}\frac{\partial V}{\partial\varphi_0}\delta\varphi\right),
    \label{eq:scalar-zz}\\
    (z,\mu):\quad &-3\nabla_\mu\psi'+3\mathcal{H}\nabla_\mu\phi
    =\frac{1}{4}\varphi_0'\nabla_\mu\delta\varphi,
    \label{eq:scalar-zmu}\\
    (\mu,\nu):\quad &\left(3\psi''-6\mathcal{H}'\phi-3\mathcal{H}\phi'
    +9\mathcal{H}\psi'-6\mathcal{H}^2\phi+\Box\phi+2\Box\psi\right)\delta^\mu_\nu
    \nonumber\\
    &\quad-\eta^{\mu\rho}\nabla_\rho\nabla_\nu\phi
    -2\eta^{\mu\rho}\nabla_\rho\nabla_\nu\psi
    \nonumber\\
    &\quad=\frac{1}{4}\left(-\varphi_0'\delta\varphi'+\phi(\varphi_0')^2
    -e^{2A}\frac{\partial V}{\partial\varphi_0}\delta\varphi\right)\delta^\mu_\nu,
    \label{eq:scalar-munu}\\
    \text{scalar}:\quad &\delta\varphi''+3\mathcal{H}\delta\varphi'
    +(4\psi'-\phi'-6\mathcal{H}\phi)\varphi_0' \nonumber\\
    &\quad-2\phi\varphi_0''+\Box\delta\varphi
    =e^{2A}\frac{\partial^2 V}{\partial\varphi_0^2}\delta\varphi.
    \label{eq:scalar-field-linearized}
\end{align}
All derivatives of $V$ in these equations are evaluated on the background $\varphi_0$.

The mixed Einstein equation~\eqref{eq:scalar-zmu} relates the scalar-field perturbation to the metric perturbations:
\be\label{eq:scl} \delta\varphi=\frac{4}{\varphi_0'}\left(-3\psi'+3\mathcal{H}\phi\right). \ee
The off-diagonal part of equation~\eqref{eq:scalar-munu} supplies a second constraint,
\be\label{eq:cons} \phi+2\psi=0. \ee
These relations leave a single independent scalar degree of freedom, which we take to be $\psi$. A complementary discussion of the scalar sector in terms of a gravitational Higgs mechanism is given in \cite{Kakushadze:2000zp}. Here we proceed directly by substituting equations~\eqref{eq:scl} and~\eqref{eq:cons} into the linearized equations, obtaining
\be\label{eq:master} \psi''+\Box\psi+\left(3\mathcal{H}-2\frac{\varphi_0''}{\varphi_0'}\right)\psi'
+\left(4\mathcal{H}'-4\mathcal{H}\frac{\varphi_0''}{\varphi_0'}\right)\psi=0. \ee
Once $\psi$ is known, $\phi$ and $\delta\varphi$ follow from the constraints.

To remove the first derivative with respect to $z$, introduce the field $F$ through
\be\label{eq:decomp} \psi(x^\mu,z)=\frac{1}{M_*^4}e^{-\frac{3}{2}A(y(z))}\varphi_0'(z)F(x^\mu,z). \ee
where the factor $M_*^{-4}$ has been restored explicitly; it is absent only when working in units with $M_*=1$. Equation~\eqref{eq:master} then becomes
\be\label{eq:F} -F''-\Box F+V_s F=0. \ee
Expanding $F$ in four-dimensional Fourier modes,
\be\label{eq:Fourier} F(x^\mu,z)=\int\frac{d^4p}{(\sqrt{2\pi})^4}\,f_p(z)e^{ip_\mu x^\mu}, \ee
with $p^2\equiv\eta^{\mu\nu}p_\mu p_\nu=-m^2$, gives
\be\label{eq:sch} -f_p''+V_s f_p=m^2 f_p, \ee
where the effective scalar potential is
\be\label{eq:effV} V_s(z)=-\frac{5}{2}\mathcal{H}'+\frac{9}{4}\mathcal{H}^2
+\mathcal{H}\frac{\varphi_0''}{\varphi_0'}-\frac{\varphi_0'''}{\varphi_0'}
+2\left(\frac{\varphi_0''}{\varphi_0'}\right)^2. \ee

\begin{figure}[htbp]
    \centering

    \begin{subfigure}{0.45\linewidth}
        \centering
        \includegraphics[width=\linewidth, height=6cm]{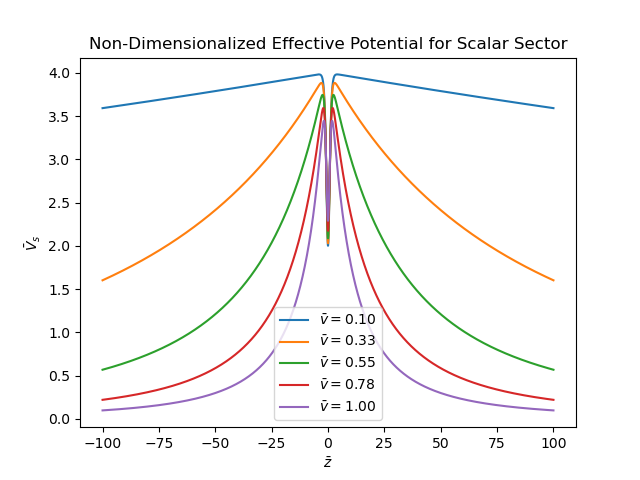}
        \caption{$\bar V_s(\bar z)$}\label{fig:scalar-effective-potential}
    \end{subfigure}
    \hspace{0.1cm}
    \begin{subfigure}{0.45\linewidth}
        \centering
        \includegraphics[width=\linewidth, height=6cm]{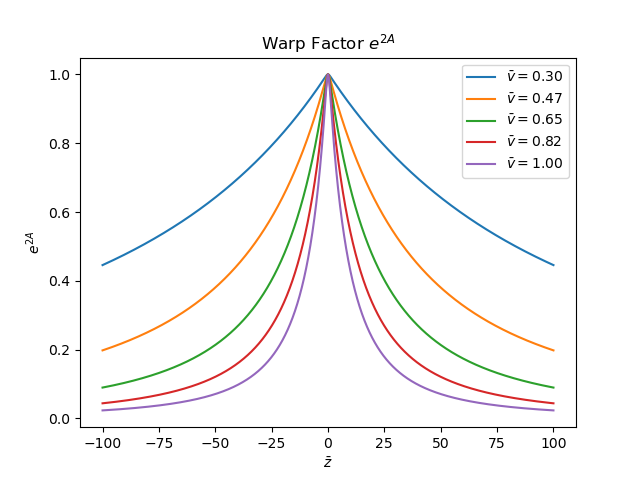}
        \caption{$A(\bar z)$}\label{fig:scalar-warp-profile}
    \end{subfigure}
    \caption{Dimensionless effective potential for scalar perturbations and the corresponding warp profile for different values of $\bar v$.}
    \label{fig:NDim}
\end{figure}

For the numerical analysis, we restore $M_*$ explicitly and introduce dimensionless variables:
\begin{align}
    \bar v&=\frac{v}{M_*^{3/2}}, & \bar\varphi&=\frac{\varphi}{M_*^{3/2}},\\
    \bar y&=ky, & \bar z&=kz,\\
    \bar V_s&=\frac{V_s}{k^2}, & \bar m&=\frac{m}{k}.
\end{align}
Substituting the kink background into equation~\eqref{eq:effV} gives the following expressions in terms of $\bar y$:
\begin{align}
    A(\bar y)&=-\frac{\bar v^2}{36}\log\!\left[\cosh^2\bar y\right]
    -\frac{\bar v^2}{72}\tanh^2\bar y, \label{eq:Absol}\\
    \bar V_s(\bar y)&=\frac{1}{1728}e^{-\frac{\bar v^2}{36}\tanh^2\bar y}
    (\cosh^2\bar y)^{-\frac{\bar v^2}{18}} \nonumber\\
    &\quad\times\Bigl[4(\bar v^2+24)(\bar v^2+72)
    -192(\bar v^2+18)\sech^2\bar y \nonumber\\
    &\hspace{2cm}-3\bar v^2(\bar v^2-104)\sech^4\bar y
    -\bar v^4\sech^6\bar y\Bigr].
    \label{eq:scalar-potential-dimensionless}
\end{align}
Here $\bar V_s(\bar y)$ denotes the potential evaluated at $\bar z(\bar y)$. Its dependence on $\bar z$ is obtained by numerically inverting the coordinate transformation~\eqref{conformal}. Figure~\ref{fig:NDim} shows the effective potential and warp profile as functions of $\bar z$ for several values of $\bar v$.

\paragraph{Search for resonances: relative probability.}
For nonzero $\bar v$, the effective potential tends to zero as $|\bar z|\to\infty$ and is positive at finite $\bar z$. It therefore admits neither normalizable tachyonic modes ($m^2<0$) nor a normalizable scalar zero mode, consistent with the discussion in \cite{Kakushadze:2000zp}. The positive-mass spectrum is continuous. Where a central well is present, however, the potential may support quasi-localized massive modes, or \textit{resonances}; the existence of a well alone does not establish their presence.

To search for resonances, we solve the dimensionless eigenvalue equation
\be\label{eq:schNonDim} -\frac{d^2 f_p}{d\bar z^2}+\bar V_s f_p=\bar m^2 f_p \ee
for $\bar V_s(0)<\bar m^2<\bar V_s^{\max}$, where $\bar V_s^{\max}$ is the barrier height. We characterize localization using the \textit{relative probability}
\be\label{eq:relProb} P(\bar m)=\frac{\displaystyle\int_{-\bar z_b}^{\bar z_b}|f_p(\bar z,\bar m^2)|^2\,d\bar z}
{\displaystyle\int_{-\bar z_{\max}}^{\bar z_{\max}}|f_p(\bar z,\bar m^2)|^2\,d\bar z}. \ee
The inner interval covers the brane region, with $\bar z_b$ of the order of the dimensionless brane thickness. The outer boundary $\bar z_{\max}$ is chosen far enough from the brane that the wavefunctions are approximately plane waves; a representative choice is $\bar z_{\max}=10\bar z_b$.

Since $\bar V_s$ is even in $\bar z$, the solutions can be separated into even and odd parity sectors. We integrate from the brane center using
\begin{align}
    f_p(0)&=1, & f_p'(0)&=0 &&\text{for even modes},\\
    f_p(0)&=0, & f_p'(0)&=1 &&\text{for odd modes},
\end{align}
where primes in these initial conditions denote derivatives with respect to $\bar z$. The nonzero initial values fix an arbitrary overall normalization, which cancels in $P(\bar m)$.

We solve equation~\eqref{eq:schNonDim} using the Numerov method \cite{Numerov:1924method,Numerov:1927note} and scan the relative probability as a function of mass. A pronounced peak indicates enhanced localization near the brane and identifies a candidate resonance. After locating the peak on a coarse mass grid, we repeat the calculation on an adaptive fine grid centered on the candidate resonance. We estimate the width in two complementary ways: from the full width at half prominence and from a Breit--Wigner fit supplemented by a smooth local background. Agreement between these estimators, together with stability under refinement of the mass grid and fitting interval, is used as a diagnostic of the extraction. The resulting full width in $\bar m$ characterizes the relative-probability peak. For an isolated resonance whose line shape is controlled by a single complex pole, this width is related to the inverse lifetime, but the relative probability is a finite-region localization diagnostic rather than a spectral density. We therefore refer to the extracted quantity as the spectral localization width and compare it separately with the time-domain energy-loss rate. Table~\ref{tab:relProb-summary} summarizes the extracted resonance masses and widths, while representative convergence tests are reported in Appendix~\ref{app:numerical-convergence}.

\begin{table}[htbp]
    \centering
    \begin{minipage}[c]{0.49\textwidth}
    \centering
    \small
    \setlength{\tabcolsep}{3pt}
    \caption{Dimensionless resonance mass and relative-probability full width at half maximum (fwhm) for different values of $\bar v$.}
    \label{tab:relProb-summary}
    \begin{tabular}{|c|c|c|}
        \hline
        \textbf{$\bar v$} & \textbf{$\bar m_{\rm res}$} & \textbf{$\bar\Gamma_{\rm fwhm}$} \\
        \hline
        1.21 & 1.75166 & $4.29\times 10^{-2}$ \\
        \hline
        1.19 & 1.75153 & $3.92\times 10^{-2}$ \\
        \hline
        1.10 & 1.74895 & $2.91\times 10^{-2}$ \\
        \hline
        0.9 & 1.74583 & $1.06\times 10^{-2}$ \\
        \hline
        0.6 & 1.74039 & $3.76\times 10^{-4}$ \\
        \hline
        0.5 & 1.73814 & $5.07\times 10^{-5}$ \\
        \hline
        0.4 & 1.73608 & $1.03\times 10^{-6}$ \\
        \hline
        0.3 & 1.73437 & $2.34\times 10^{-10}$ \\
        \hline
    \end{tabular}
    \end{minipage}\hfill
    \begin{minipage}[c]{0.49\textwidth}
        \centering
        \includegraphics[width=\linewidth, height=6cm]{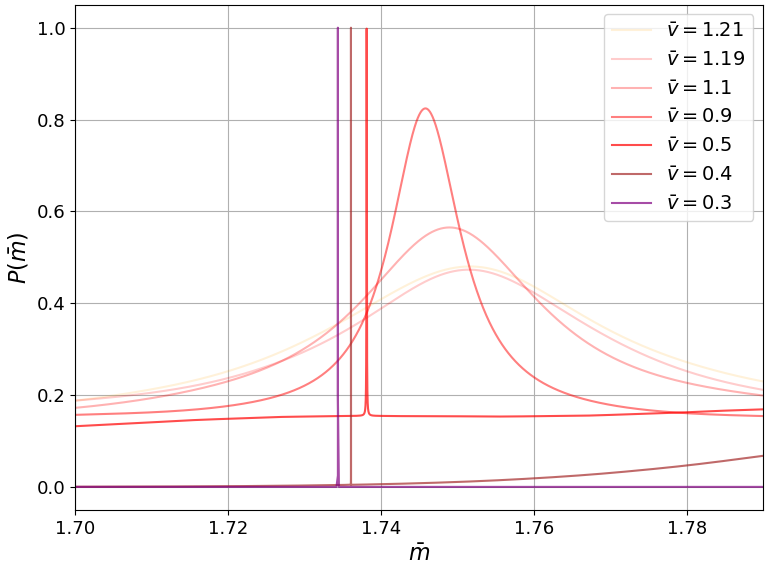}
        \captionof{figure}{Relative-probability peaks for different values of $\bar v$. As $\bar v$ decreases, the resonance becomes narrower and its mass approaches $\bar m^2=3$.}
        \label{fig:Spec}
    \end{minipage}
\end{table}

\noindent Figure~\ref{fig:Spec} shows a resonance peak that becomes narrower and approaches $\bar m^2=3$ as $\bar v$ decreases. The width becomes very small but remains nonzero at $\bar v=0.3$: Table~\ref{tab:relProb-summary} gives $\bar\Gamma_{\mathrm{fwhm}}=2.34\times10^{-10}$. The approach to $\bar m^2=3$, together with the narrowing of the peak, motivates the comparison with the flat-space shape mode in the next section.

\section{The scalar resonant mode: a relic of the shape mode}\label{sec:d}
The coupled scalar--gravity system has no normalizable scalar bound modes, in contrast to the flat-space $\varphi^4$ kink, which supports a translational zero mode and a shape mode. As discussed in Section~\ref{sec:c}, the scalar-field perturbation is constrained by the metric perturbations, so the flat-space spectrum cannot be carried over unchanged to the gravitating system. We now focus on the massive resonance and examine its relation to the kink's shape mode.

\begin{figure}[htbp]
    \centering
    \begin{subfigure}{0.45\linewidth}
        \centering
        \includegraphics[width=\linewidth, height=6cm]{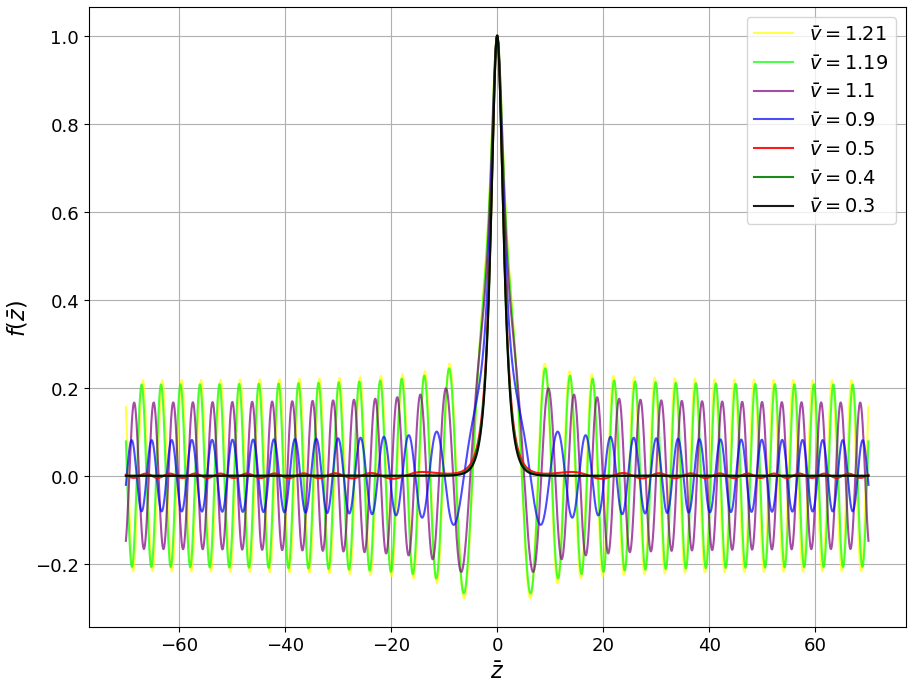}
        \caption{}\label{fig:shape-resonant-profile}
    \end{subfigure}
    \hspace{0.1cm}
    \begin{subfigure}{0.45\linewidth}
        \centering
        \includegraphics[width=\linewidth, height=6cm]{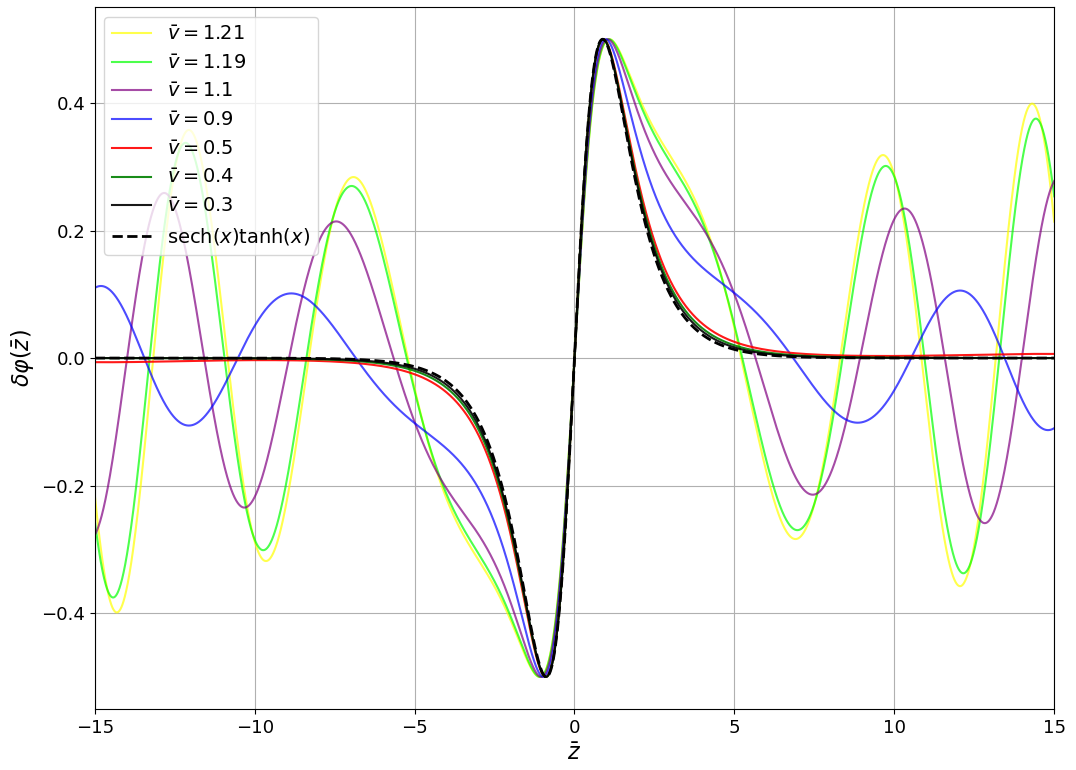}
        \caption{}\label{fig:shape-scalar-profile}
    \end{subfigure}
    \caption{Resonant mode profiles for different values of $\bar v$. The oscillatory tails become less prominent as $\bar v$ decreases. In panel (b), the near-brane profile of the scalar-field perturbation $\delta\varphi_p$ approaches the flat-space shape-mode profile, up to an overall normalization.}
    \label{fig:shape-resonant-profiles}
\end{figure}

\noindent For each resonance mass in Table~\ref{tab:relProb-summary}, we compute the corresponding profile $f_p(\bar z)$. The metric perturbation $\psi_p$ and scalar-field perturbation $\delta\varphi_p$ are then reconstructed using equations~\eqref{eq:decomp}, \eqref{eq:cons}, and~\eqref{eq:scl}. Figure~\ref{fig:shape-resonant-profiles} shows the resulting resonant profiles for several values of $\bar v$. As $\bar v$ decreases, the near-brane profile of $\delta\varphi_p$ approaches the flat-space shape-mode profile described in Section~\ref{sec:a}, up to an overall normalization.

Two observations therefore support the identification of this resonance as a remnant of the shape mode: its mass approaches $\bar m^2=3$, and its scalar-field profile approaches that of the flat-space bound state. The gravitationally modified potential allows the excitation to leak into the bulk, so it appears as a quasi-localized resonance rather than a normalizable bound mode. The increasingly sharp resonance at smaller $\bar v$ is consistent with recovery of the bound shape mode in the flat-space limit.

\begin{figure}[htbp]
    \centering
    \begin{subfigure}{0.45\linewidth}
        \centering
        \includegraphics[width=\linewidth, height=6cm]{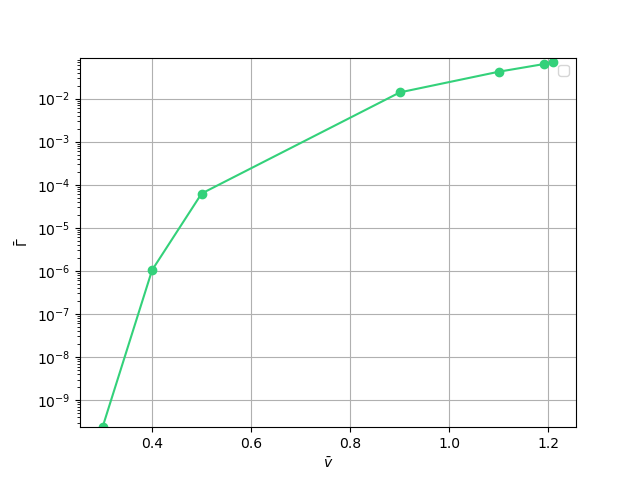}
        \caption{}\label{fig:shape-width}
    \end{subfigure}
    \hspace{0.1cm}
    \begin{subfigure}{0.45\linewidth}
        \centering
        \includegraphics[width=\linewidth, height=6cm]{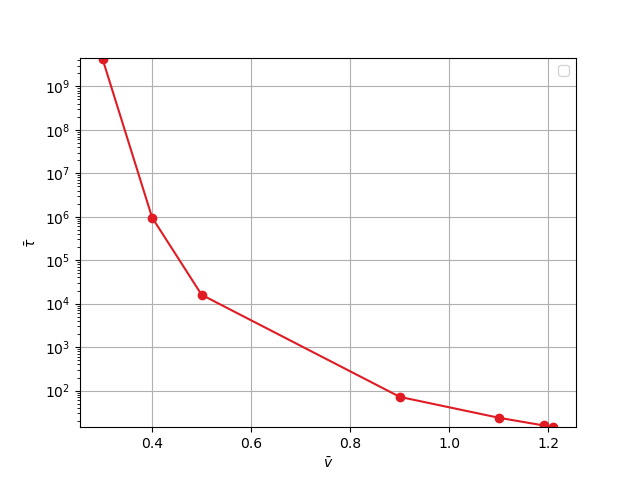}
        \caption{}\label{fig:shape-lifetime}
    \end{subfigure}
    \caption{(a) Dimensionless relative-probability width $\bar\Gamma_{\rm spec}$ and (b) its inverse, $\bar\tau_{\rm spec}=1/\bar\Gamma_{\rm spec}$, as functions of $\bar v$. The inverse width is the spectral localization timescale.}
    \label{fig:shape-width-lifetime}
\end{figure}

\noindent The corresponding widths and inverse-width scales are shown in Figure~\ref{fig:shape-width-lifetime}. As $\bar v$ decreases, the relative-probability width $\bar\Gamma_{\rm spec}$ decreases and $\bar\tau_{\rm spec}=1/\bar\Gamma_{\rm spec}$ increases. This trend is consistent with an excitation that remains concentrated near the brane for progressively longer times. The mass, profile, and narrowing of the localization peak provide complementary evidence for the shape-mode interpretation. In the next section, we evolve the resonance directly and extract an independent localized-energy leakage rate.

\section{Numerical time evolution and characterization of the shape-mode resonance}\label{sec:f}
We now evolve the resonant profile identified in Section~\ref{sec:d} to examine its time dependence and compare its oscillation frequency with the mass extracted from the static analysis. We use a method-of-lines implementation adapted to the kink-brane background, following the approach of Section~3.2 of \cite{Tan:2022uex}.

\subsection{Evolution equation and first-order reduction}

For a spatial Fourier mode along the brane, let $f(t,z)$ denote the time-dependent amplitude of the field $F$ introduced in equation~\eqref{eq:decomp}. In this section, we suppress bars on the dimensionless coordinates $\bar t$, $\bar z$, and potential $\bar V_s$. We denote the dimensionless three-momentum magnitude by $q=|\vec p|$. The evolution equation is
\begin{equation}
  -\partial_t^2 f(t,z)+\partial_z^2 f(t,z)-V_s(z)f(t,z)-q^2 f(t,z)=0.
  \label{eq:evolution}
\end{equation}
We restrict to $q=0$, so the perturbation is spatially homogeneous along the brane but retains its dependence on time and the extra dimension, as in \cite{Tan:2022uex}.

Introducing the auxiliary field $\Pi\equiv\partial_t f$ gives the first-order system
\begin{align}
  \partial_t f&=\Pi,\\
  \partial_t\Pi&=\partial_z^2 f-V_s(z)f.
\end{align}
We discretize the spatial operators on a uniform grid $z_i=-z_{\max}+i\Delta z$, with $i=0,\ldots,N-1$ and $\Delta z=2z_{\max}/(N-1)$. Fourth-order finite-difference stencils are centered in the interior and one-sided near the boundaries. For example,
\begin{equation}
  \left.\partial_z f\right|_i\approx
  \frac{-f_{i+2}+8f_{i+1}-8f_{i-1}+f_{i-2}}{12\Delta z},
  \qquad 2\leq i\leq N-3.
\end{equation}
Fourth-order one-sided expressions are used at $i=0,1,N-2,N-1$, and the second derivative is discretized to the same order.

The resulting system of ordinary differential equations,
\begin{equation}
  \frac{d}{dt}\begin{pmatrix}f\\\Pi\end{pmatrix}
  =\mathcal{L}\begin{pmatrix}f\\\Pi\end{pmatrix},
\end{equation}
is integrated with the adaptive Runge--Kutta method RK23. The integrator selects the time step subject to the Courant-type ceiling $\Delta t_{\max}=0.2\Delta z$. We use $z_{\max}=70$, which places the numerical boundaries well beyond the potential barriers for masses near the resonance, together with a grid of $N=2800$ points.

\subsection{Boundary treatment: sponge layer and dissipation}

Outgoing radiation must be allowed to leave the region of interest without significant reflection from the finite-domain boundaries. The approach of \cite{Tan:2022uex} uses dissipative boundary conditions. In our discretization, a direct characteristic-penalty implementation was found to be linearly unstable. We therefore use a smooth absorbing \emph{sponge layer}, adding a damping term to the momentum equation:
\begin{equation}
  \partial_t\Pi=\partial_z^2 f-V_s(z)f-\gamma(z)\Pi,
\end{equation}
where
\begin{equation}
  \gamma(z)=\gamma_0\left[\max\!\left(0,1-\frac{z_{\max}-|z|}{L_{\rm sp}}\right)\right]^2.
\end{equation}
The damping vanishes in the interior and increases to $\gamma_0$ over a layer of width $L_{\rm sp}$ near each boundary. Its purpose is to attenuate outgoing waves before they reach the outer boundary, thereby reducing reflected contamination. The sponge is an approximate absorbing treatment rather than an exactly reflection-free boundary condition. To suppress grid-scale oscillations, including noise generated by the one-sided boundary stencils, we additionally apply Kreiss--Oliger dissipation using a sixth-difference stencil. Its contribution to the field evolution is
\begin{equation}
  \left.\frac{df_i}{dt}\right|_{\rm KO}
  =\frac{\epsilon}{64\Delta z}
  \left(f_{i+3}-6f_{i+2}+15f_{i+1}-20f_i+15f_{i-1}-6f_{i-2}+f_{i-3}\right).
\end{equation}
The same operator is applied to $\Pi_i$ wherever the stencil fits within the grid, with $\epsilon\sim0.3$.

\subsection{Initial data}

We use the real resonant profile $f_p(z)$ obtained by the Numerov and relative-probability procedure of Section~\ref{sec:c} as the initial field configuration:
\begin{equation}
  f(0,z)=f_p(z),\qquad \Pi(0,z)=0.
\end{equation}
Because the real-frequency scattering profile has oscillatory tails, it is smoothly truncated before being used on the finite numerical domain. We multiply it by an even cosine taper $W(z)$ that equals unity for $|z|\leq z_{\rm tap}^{\rm in}$, decreases continuously to zero for $z_{\rm tap}^{\rm in}<|z|<z_{\rm tap}^{\rm out}$, and vanishes beyond $z_{\rm tap}^{\rm out}$. Explicitly,
\begin{equation}
 W(z)=
 \begin{cases}
 1, & |z|\leq z_{\rm tap}^{\rm in},\\[2mm]
 \dfrac{1}{2}\left[1+\cos\!\left(\pi\dfrac{|z|-z_{\rm tap}^{\rm in}}
 {z_{\rm tap}^{\rm out}-z_{\rm tap}^{\rm in}}\right)\right],
 & z_{\rm tap}^{\rm in}<|z|<z_{\rm tap}^{\rm out},\\[3mm]
 0, & |z|\geq z_{\rm tap}^{\rm out}.
 \end{cases}
\end{equation}
Thus the evolved initial profile is $f_w(0,z)=W(z)f_p(z)$. The taper begins outside the outer classical turning point and outside the region used to measure the localized energy. Its position is varied in the convergence study so that the extracted decay rate is not determined by the artificial truncation. The initial signal may contain a transient arising from radiative components of the initial data and their adjustment to the numerical boundary treatment. Thus the late-time analysis concerns the evolution after this initial adjustment, when the resonance dominates the signal.

\subsection{Time series and frequency spectra}

\begin{figure}[htbp]
    \centering
    \captionsetup[subfigure]{font=small}
    \begin{subfigure}{0.32\linewidth}
        \centering
        \includegraphics[width=\linewidth,height=3.4cm,keepaspectratio]{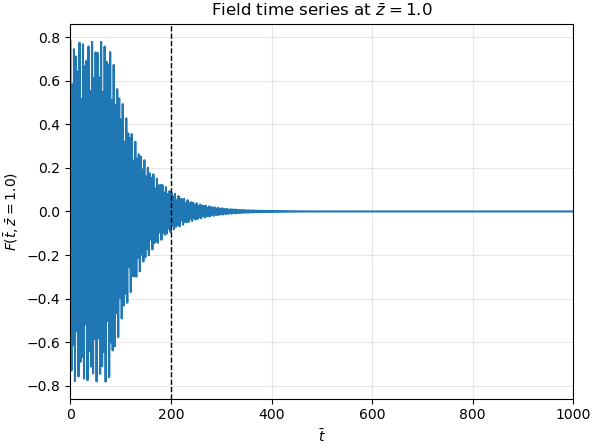}
        \caption{$\bar v=1.21$}\label{fig:evolution-time-121}
    \end{subfigure}%
    \hfill
    \begin{subfigure}{0.32\linewidth}
        \centering
        \includegraphics[width=\linewidth,height=3.4cm,keepaspectratio]{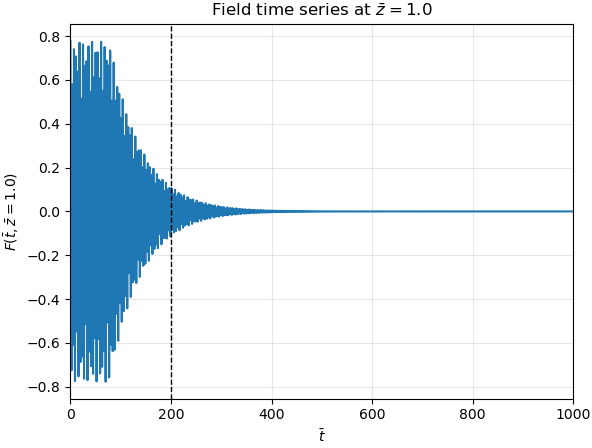}
        \caption{$\bar v=1.19$}\label{fig:evolution-time-119}
    \end{subfigure}%
    \hfill
    \begin{subfigure}{0.32\linewidth}
        \centering
        \includegraphics[width=\linewidth,height=3.4cm,keepaspectratio]{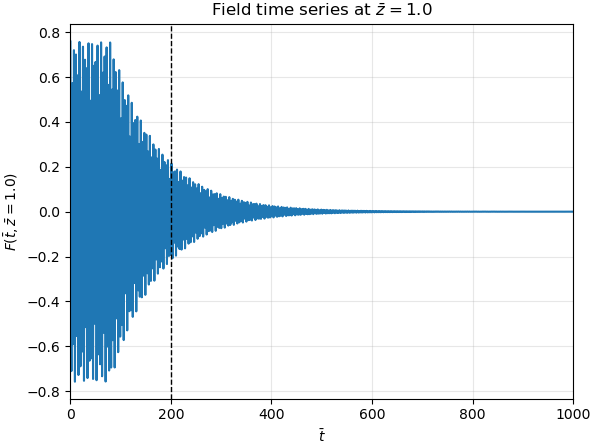}
        \caption{$\bar v=1.1$}\label{fig:evolution-time-11}
    \end{subfigure}
    \par\medskip
    \begin{subfigure}{0.32\linewidth}
        \centering
        \includegraphics[width=\linewidth,height=3.4cm,keepaspectratio]{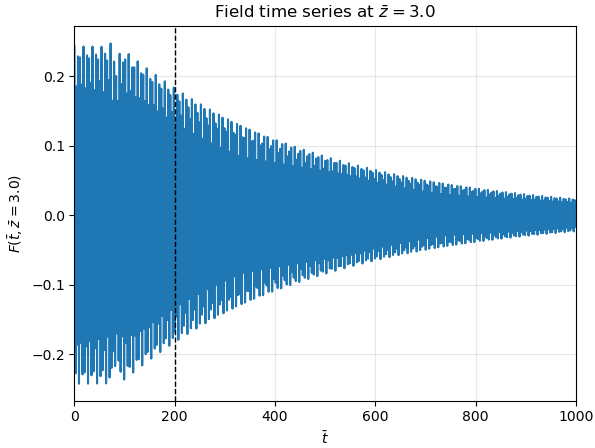}
        \caption{$\bar v=0.9$}\label{fig:evolution-time-09}
    \end{subfigure}%
    \hspace{0.02\linewidth}
    \begin{subfigure}{0.32\linewidth}
        \centering
        \includegraphics[width=\linewidth,height=3.4cm,keepaspectratio]{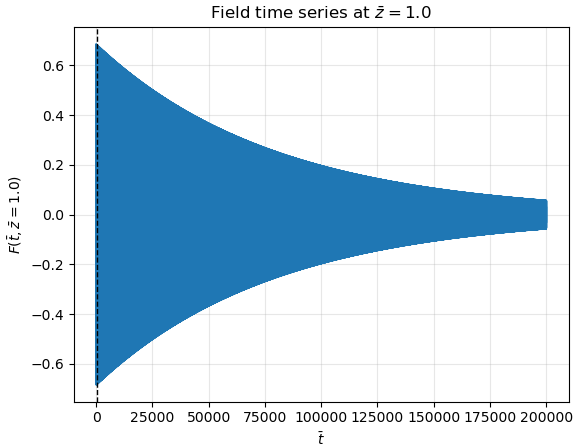}
        \caption{$\bar v=0.6$}\label{fig:evolution-time-06}
    \end{subfigure}
    \caption{Field time series at the extraction points indicated in the panels, for $\bar v=1.21,1.19,1.1,0.9,0.6$ in panels (a)-(e), respectively. The oscillation envelope decays after an initial transient. The time ranges differ between panels to display the increasingly slow decay at smaller $\bar v$.}
    \label{fig:evolution-time-series}
\end{figure}

We record the field at a fixed extraction point for each run; the coordinates are indicated in the plot titles. Figure~\ref{fig:evolution-time-series} shows the time series for five values of $\bar v$. Following the initial transient, the signal exhibits damped oscillations whose envelope is consistent with exponential decay over the displayed late-time interval. The slower decay at smaller $\bar v$ is consistent with the narrowing of the spectral resonance discussed in Section~\ref{sec:d}.

\begin{figure}[htbp]
    \centering
    \captionsetup[subfigure]{font=small}
    \begin{subfigure}{0.32\linewidth}
        \centering
        \includegraphics[width=\linewidth,height=3.4cm,keepaspectratio]{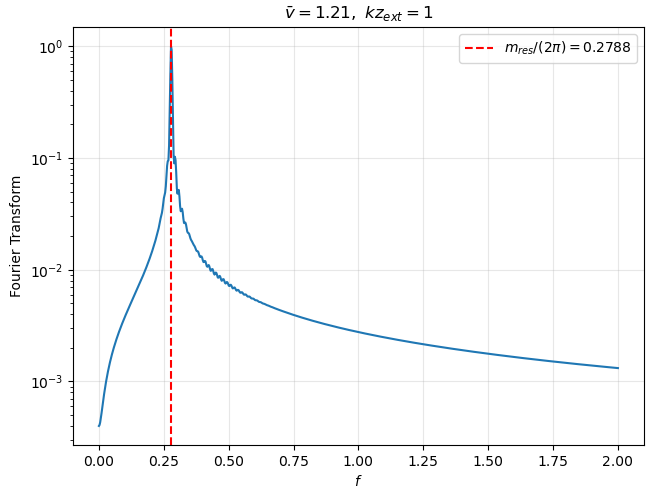}
        \caption{$\bar v=1.21$}\label{fig:evolution-spectrum-121}
    \end{subfigure}%
    \hfill
    \begin{subfigure}{0.32\linewidth}
        \centering
        \includegraphics[width=\linewidth,height=3.4cm,keepaspectratio]{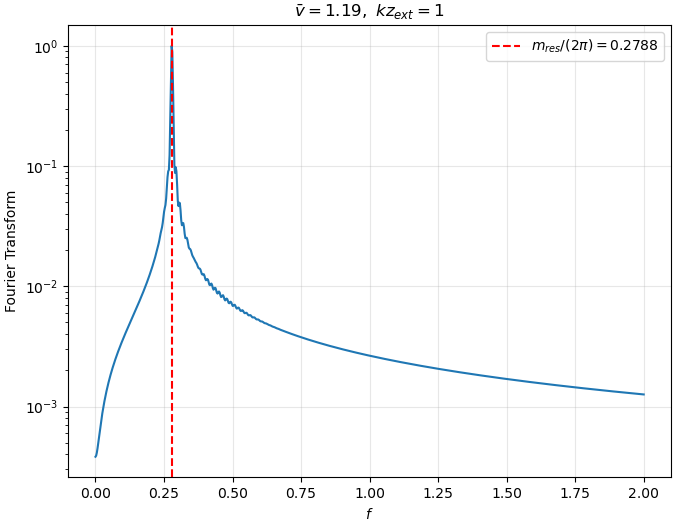}
        \caption{$\bar v=1.19$}\label{fig:evolution-spectrum-119}
    \end{subfigure}%
    \hfill
    \begin{subfigure}{0.32\linewidth}
        \centering
        \includegraphics[width=\linewidth,height=3.4cm,keepaspectratio]{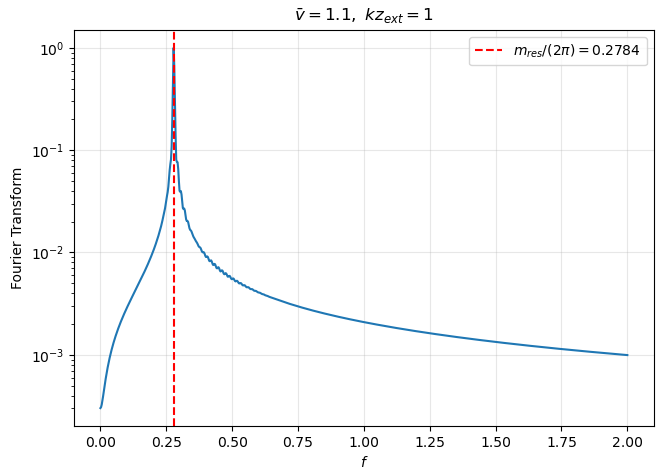}
        \caption{$\bar v=1.1$}\label{fig:evolution-spectrum-11}
    \end{subfigure}
    \par\medskip
    \begin{subfigure}{0.32\linewidth}
        \centering
        \includegraphics[width=\linewidth,height=3.4cm,keepaspectratio]{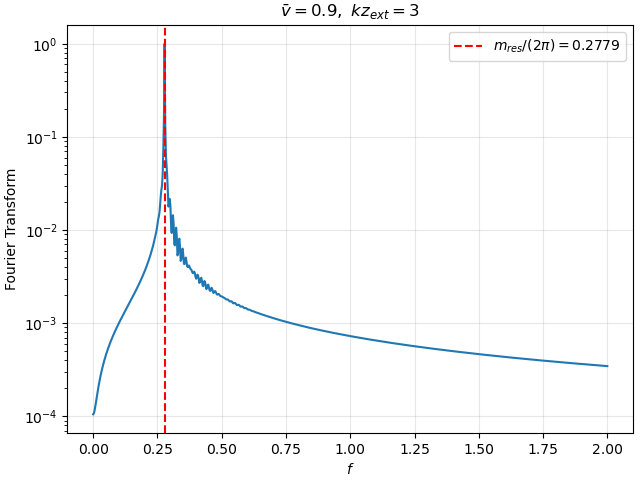}
        \caption{$\bar v=0.9$}\label{fig:evolution-spectrum-09}
    \end{subfigure}%
    \hspace{0.02\linewidth}
    \begin{subfigure}{0.32\linewidth}
        \centering
        \includegraphics[width=\linewidth,height=3.4cm,keepaspectratio]{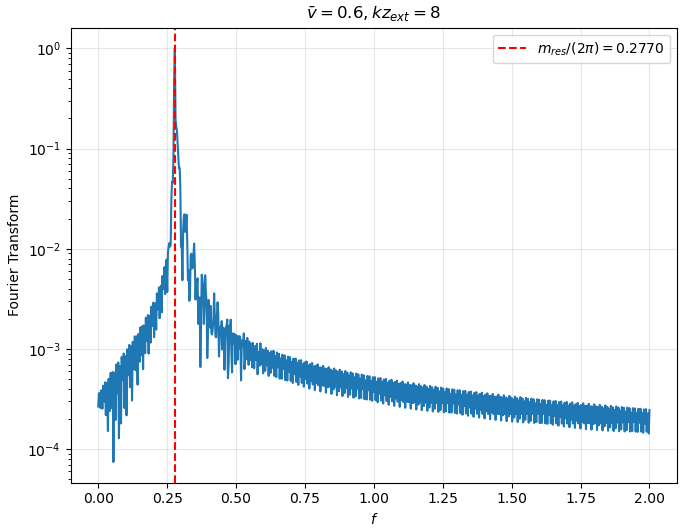}
        \caption{$\bar v=0.6$}\label{fig:evolution-spectrum-06}
    \end{subfigure}
    \caption{Discrete Fourier spectra of the extracted signals for $\bar v=1.21,1.19,1.1,0.9,0.6$ in panels (a)-(e), respectively. The dominant peaks lie near the frequencies $\bar m_{\rm res}/(2\pi)$ inferred from the static resonance masses.}
    \label{fig:evolution-frequency-spectra}
\end{figure}

The discrete Fourier spectra of the extracted signals are shown in Figure~\ref{fig:evolution-frequency-spectra}. Each spectrum has a dominant peak near the frequency predicted by the static resonance mass. Since the Fourier frequency counts cycles per unit dimensionless time, the expected peak for $q=0$ is at $\nu=\bar m_{\rm res}/(2\pi)$. The symbol $f$ used on the plot axes denotes this Fourier frequency and should not be confused with the evolved field $f(t,z)$. Agreement between the spectral peak and the static mass supports the resonance interpretation when considered together with localization and time-domain decay.

The numerical parameters are not inferred from a single run. We vary the grid spacing, localized-energy radius, taper interval, fit window, sponge strength and width, and dissipation coefficient. Representative convergence tests for a moderately broad and a narrow resonance are summarized in Appendix \ref{app:numerical-convergence}.

\section{WKB analysis of the resonance parameters}\label{sec:add1}
The preceding numerical analysis shows that the scalar resonance approaches the flat-space shape mode as the gravitational coupling is reduced. An analytic treatment of the exact effective potential is difficult because the transformation from $y$ to the conformal coordinate $z$ is not available in closed form. In this section, we therefore use a four-turning-point WKB analysis of a symmetric model potential that retains the central well and the two barriers of $V_s(z)$. This approximation provides a simple analytic description of the resonance mass and explains the exponential suppression of its width as the barriers become less penetrable.

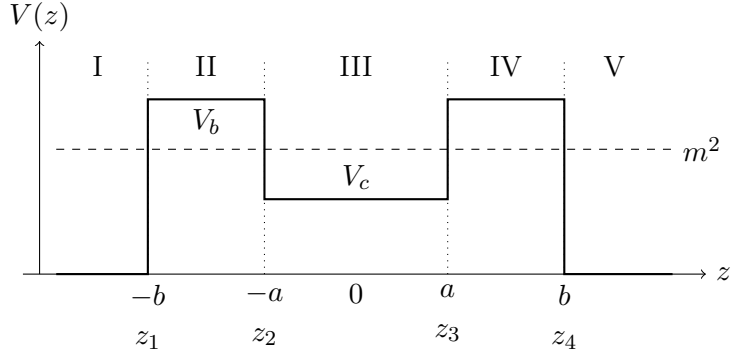
\begin{figure}[htbp]
    \centering
    \begin{tikzpicture}[xscale=1.1,yscale=1.1]
        \draw[->] (-4,0) -- (4.2,0) node[right] {$z$};
        \draw[->] (-3.8,0) -- (-3.8,2.8) node[above] {$V(z)$};
        \draw[thick] (-3.6,0) -- (-2.5,0) -- (-2.5,2.1)
            -- (-1.1,2.1) -- (-1.1,0.9) -- (1.1,0.9)
            -- (1.1,2.1) -- (2.5,2.1) -- (2.5,0) -- (3.8,0);
        \draw[dashed] (-3.6,1.5) -- (3.8,1.5) node[right] {$m^2$};
        \draw[dotted] (-2.5,0) -- (-2.5,2.6);
        \draw[dotted] (-1.1,0) -- (-1.1,2.6);
        \draw[dotted] (1.1,0) -- (1.1,2.6);
        \draw[dotted] (2.5,0) -- (2.5,2.6);
        \node at (-3.1,2.5) {I};
        \node at (-1.8,2.5) {II};
        \node at (0,2.5) {III};
        \node at (1.8,2.5) {IV};
        \node at (3.1,2.5) {V};
        \node[below] at (-1.8,2.1) {$V_b$};
        \node[above] at (0,0.9) {$V_c$};
        \node[below,align=center] at (-2.5,0) {$-b$\\$z_1$};
        \node[below,align=center] at (-1.1,0) {$-a$\\$z_2$};
        \node[below] at (0,0) {$0$};
        \node[below,align=center] at (1.1,0) {$a$\\$z_3$};
        \node[below,align=center] at (2.5,0) {$b$\\$z_4$};
    \end{tikzpicture}
    \caption{Schematic double-barrier potential and the five WKB regions. The rectangular profile illustrates the region labels.}
    \label{fig:wkb-regions}
\end{figure}

We begin with the Schr\"{o}dinger equation for the scalar mode, approximating $V_s(z)$ as $V(z)$ (see Figure \ref{fig:wkb-regions}) and denoting the mode function by $\Psi(z)$ throughout this section:
\begin{align}
    -\Psi''+V(z)\Psi&=m^2\Psi, \\
    \Psi''+\mathcal{Q}(z)\Psi&=0, \qquad \mathcal{Q}(z)\equiv m^2-V(z).
    \label{eq:wkb-schrodinger}
\end{align}
The local wave numbers in the classically allowed and forbidden regions are, respectively,
\begin{equation}\label{eq:wkb-wave-numbers}
    p(z)\equiv\sqrt{m^2-V(z)}, \qquad
    \kappa(z)\equiv\sqrt{V(z)-m^2}.
\end{equation}
Here $p(z)$ denotes the local wave number, while $k$ continues to denote the constant inverse kink width.

For a symmetric double-barrier potential, we label the four turning points by
\begin{equation}
    z_1=-b, \qquad z_2=-a, \qquad z_3=a, \qquad z_4=b,
    \qquad 0<a<b.
\end{equation}
These divide the problem into five regions: the allowed regions I, III and V, and the forbidden regions II and IV, as illustrated in Figure~\ref{fig:wkb-regions}.

We formulate the WKB problem in terms of real, delta-function-normalized continuum scattering states. By parity symmetry, it is sufficient to perform the matching on the right half-line and impose the appropriate condition at $z=0$.
The WKB solutions in the three allowed regions can be written as
\begin{align}
    \Psi_{\mathrm{I}}(z)
    &=\frac{1}{\sqrt{p(z)}}\left[
        A_{\mathrm{I}}e^{i\int_z^{z_1}p(z')\,\d z'}
        +B_{\mathrm{I}}e^{-i\int_z^{z_1}p(z')\,\d z'}\right],\\
    \Psi_{\mathrm{III}}(z)
    &=\frac{1}{\sqrt{p(z)}}\left[
        A_{\mathrm{III}}e^{i\int_{z_2}^{z}p(z')\,\d z'}
        +B_{\mathrm{III}}e^{-i\int_{z_2}^{z}p(z')\,\d z'}\right],\\
    \Psi_{\mathrm{V}}(z)
    &=\frac{1}{\sqrt{p(z)}}\left[
        A_{\mathrm{V}}e^{i\int_{z_4}^{z}p(z')\,\d z'}
        +B_{\mathrm{V}}e^{-i\int_{z_4}^{z}p(z')\,\d z'}\right].
    \label{eq:wkb-allowed}
\end{align}
In the two forbidden regions, the corresponding solutions are
\begin{align}
    \Psi_{\mathrm{II}}(z)
    &=\frac{1}{\sqrt{\kappa(z)}}\left[
        A_{\mathrm{II}}e^{\int_{z_1}^{z}\kappa(z')\,\d z'}
        +B_{\mathrm{II}}e^{-\int_{z_1}^{z}\kappa(z')\,\d z'}\right],\\
    \Psi_{\mathrm{IV}}(z)
    &=\frac{1}{\sqrt{\kappa(z)}}\left[
        A_{\mathrm{IV}}e^{\int_z^{z_4}\kappa(z')\,\d z'}
        +B_{\mathrm{IV}}e^{-\int_z^{z_4}\kappa(z')\,\d z'}\right].
    \label{eq:wkb-forbidden}
\end{align}

\subsection{Even mode in the central well}

The resonance of interest belongs to the even sector of the master variable $\Psi$. Accordingly, $\Psi'_{\mathrm{III}}(0)=0$. To impose this condition, define
\begin{equation}
    \theta(z)\equiv\int_{z_2}^{z}p(z')\,\d z', \qquad
    \theta_0\equiv\theta(0)=\int_{z_2}^{0}p(z')\,\d z', \qquad
    \theta'(z)=p(z).
\end{equation}
Rewriting the central solution in a trigonometric basis gives
\begin{equation}
    \Psi_{\mathrm{III}}(z)=\frac{1}{\sqrt{p(z)}}
    \left[C_{\mathrm{III}}\cos\theta(z)+D_{\mathrm{III}}\sin\theta(z)\right].
\end{equation}
Its derivative is
\begin{align}
    \Psi'_{\mathrm{III}}(z)
    &=-\frac{p'}{2p^{3/2}}
    \left[C_{\mathrm{III}}\cos\theta+D_{\mathrm{III}}\sin\theta\right]
    \nonumber\\
    &\quad+\frac{\theta'}{\sqrt{p}}
    \left[-C_{\mathrm{III}}\sin\theta+D_{\mathrm{III}}\cos\theta\right].
\end{align}
Since $V'(0)=0$, we have $p'(0)=0$. Thus the even-mode condition becomes
\begin{equation}
    \sqrt{p(0)}\left[D_{\mathrm{III}}\cos\theta_0
    -C_{\mathrm{III}}\sin\theta_0\right]=0.
\end{equation}
For $p(0)\ne0$, this yields $D_{\mathrm{III}}=C_{\mathrm{III}}\tan\theta_0$, and hence
\begin{align}
    \Psi_{\mathrm{III}}(z)
    &=\frac{C_{\mathrm{III}}}{\sqrt{p(z)}}
    \left[\cos\theta+\tan\theta_0\sin\theta\right]\\
    &=\frac{C_{\mathrm{III}}}{\sqrt{p(z)}\cos\theta_0}
    \cos(\theta-\theta_0)\\
    &=\frac{\widetilde C_{\mathrm{III}}}{\sqrt{p(z)}}
    \cos\left(\int_0^z p(z')\,\d z'\right),
    \label{eq:wkb-even-mode}
\end{align}
where $\widetilde C_{\mathrm{III}}\equiv C_{\mathrm{III}}/\cos\theta_0$ is an overall amplitude. Define the half-well phase and the phase measured from the right inner turning point by
\begin{equation}\label{eq:wkb-well-phases}
    \Theta\equiv\int_0^{z_3}p(z')\,\d z', \qquad
    \widetilde\theta(z)\equiv\int_z^{z_3}p(z')\,\d z'.
\end{equation}
Since $\int_0^z p(z')\,\d z'=\Theta-\widetilde\theta(z)$, the central solution takes the form
\begin{align}
    \Psi_{\mathrm{III}}(z)
    &=\frac{\widetilde C_{\mathrm{III}}}{\sqrt{p(z)}}
    \cos\left(\Theta-\widetilde\theta(z)\right)\\
    &=\frac{\widetilde C_{\mathrm{III}}}{\sqrt{p(z)}}
    \left[\cos\Theta\cos\widetilde\theta
    +\sin\Theta\sin\widetilde\theta\right].
    \label{eq:wkb-central-matching}
\end{align}

\subsection{Matching regions III and IV}

Define
\begin{equation}\label{eq:wkb-barrier-action}
    S\equiv\int_{z_3}^{z_4}\kappa(z')\,\d z', \qquad
    \eta(z)\equiv\int_{z_3}^{z}\kappa(z')\,\d z'.
\end{equation}
Then $\int_z^{z_4}\kappa(z')\,\d z'=S-\eta(z)$, and the region-IV solution becomes
\begin{equation}\label{eq:wkb-barrier-inner}
    \Psi_{\mathrm{IV}}(z)=\frac{1}{\sqrt{\kappa(z)}}
    \left[A_{\mathrm{IV}}e^S e^{-\eta(z)}+B_{\mathrm{IV}}e^{-S}e^{\eta(z)}\right].
\end{equation}
The first and second terms are, respectively, decreasing and increasing as one moves from $z_3$ toward $z_4$. The connection formulas at the inner turning point are
\begin{align}
    \frac{e^{-\eta(z)}}{\sqrt{\kappa(z)}}
    &\longleftrightarrow\frac{2}{\sqrt{p(z)}}
    \cos\left(\widetilde\theta(z)-\frac{\pi}{4}\right),\\
    \frac{e^{\eta(z)}}{\sqrt{\kappa(z)}}
    &\longleftrightarrow-\frac{1}{\sqrt{p(z)}}
    \sin\left(\widetilde\theta(z)-\frac{\pi}{4}\right).
    \label{eq:wkb-inner-connection}
\end{align}
Consequently, the continuation of the barrier solution into region III is
\begin{align}
    \Psi_{\mathrm{IV}\to\mathrm{III}}(z)
    &=\frac{1}{\sqrt{p(z)}}\left[
    2A_{\mathrm{IV}}e^S\cos\left(\widetilde\theta-\frac{\pi}{4}\right)
    -B_{\mathrm{IV}}e^{-S}\sin\left(\widetilde\theta-\frac{\pi}{4}\right)\right]\\
    &=\frac{1}{\sqrt{2p(z)}}\left[
    (2A_{\mathrm{IV}}e^S+B_{\mathrm{IV}}e^{-S})\cos\widetilde\theta
    +(2A_{\mathrm{IV}}e^S-B_{\mathrm{IV}}e^{-S})\sin\widetilde\theta\right].
\end{align}
Matching the coefficients with equation~\eqref{eq:wkb-central-matching} gives
\begin{align}
    \widetilde C_{\mathrm{III}}\cos\Theta
    &=\frac{1}{\sqrt{2}}\left(2A_{\mathrm{IV}}e^S+B_{\mathrm{IV}}e^{-S}\right),\\
    \widetilde C_{\mathrm{III}}\sin\Theta
    &=\frac{1}{\sqrt{2}}\left(2A_{\mathrm{IV}}e^S-B_{\mathrm{IV}}e^{-S}\right).
\end{align}
Solving for the barrier coefficients, we obtain
\begin{align}
    A_{\mathrm{IV}}&=\frac{\widetilde C_{\mathrm{III}}}{2\sqrt{2}}
    e^{-S}\left(\cos\Theta+\sin\Theta\right),\\
    B_{\mathrm{IV}}&=\frac{\widetilde C_{\mathrm{III}}}{\sqrt{2}}
    e^S\left(\cos\Theta-\sin\Theta\right).
    \label{eq:wkb-barrier-coefficients}
\end{align}

\subsection{Matching regions IV and V}

At the outer turning point $z_4$, introduce
\begin{equation}
    \xi(z)\equiv\int_{z_4}^{z}p(z')\,\d z', \qquad
    \zeta(z)\equiv\int_z^{z_4}\kappa(z')\,\d z', \qquad
    \zeta(z_4)=0.
\end{equation}
The barrier solution is then
\begin{equation}
    \Psi_{\mathrm{IV}}(z)=\frac{1}{\sqrt{\kappa(z)}}
    \left[A_{\mathrm{IV}}e^{\zeta(z)}+B_{\mathrm{IV}}e^{-\zeta(z)}\right].
\end{equation}
Using the connection formulas
\begin{align}
    \frac{e^{-\zeta(z)}}{\sqrt{\kappa(z)}}
    &\longleftrightarrow\frac{2}{\sqrt{p(z)}}
    \cos\left(\xi(z)-\frac{\pi}{4}\right),\\
    \frac{e^{\zeta(z)}}{\sqrt{\kappa(z)}}
    &\longleftrightarrow-\frac{1}{\sqrt{p(z)}}
    \sin\left(\xi(z)-\frac{\pi}{4}\right),
    \label{eq:wkb-outer-connection}
\end{align}
and defining $\Sigma(m)=\Theta(m)-\pi/4$ we find
\begin{align}
    \Psi_{\mathrm{V}}(z)
    &=\frac{1}{\sqrt{p(z)}}\left[
    -A_{\mathrm{IV}}\sin\left(\xi-\frac{\pi}{4}\right)
    +2B_{\mathrm{IV}}\cos\left(\xi-\frac{\pi}{4}\right)\right]\\
    &=\frac{\widetilde C_{\mathrm{III}}}{\sqrt{p(z)}}
    \left[-2e^{S(m)}\sin{\Sigma(m)}\cos\left(\xi-\frac{\pi}{4}\right)-\frac{1}{2}e^{-S(m)}\cos{\Sigma(m)}\sin\left(\xi-\frac{\pi}{4}\right)\right]
    \label{eq:wkb-travelling-waves}
\end{align}

\subsection{Resonance condition, mass, and width}
Delta-function normalization of the wavefunction $\Psi_{\mathrm{V}}(z)$ at infinity fixes the coefficient $\widetilde C_{\mathrm{III}}$ in equation~\eqref{eq:wkb-travelling-waves}. As $z\to\infty$, the wavefunction takes the form $\pi^{-1/2}\cos(mz+\Phi)$, with $\int \Psi_m\Psi_{m'}\,dz=\delta(m-m')$. The normalization constant is therefore

\begin{align}
|\widetilde C_{\mathrm{III}}|^2=\frac{m}{\pi}\frac{1}{\frac{1}{4}e^{-2S(m)}\cos^2{\Sigma(m)}+4e^{2S(m)}\sin^2\Sigma(m)}\\
\therefore |\Psi_m(0)|^2=\frac{m}{\pi p(0)}\frac{1}{\frac{1}{4}e^{-2S(m)}\cos^2{\Sigma(m)}+4e^{2S(m)}\sin^2\Sigma(m)}
\end{align}
For $S(m)\gg1$ and away from resonance, the term proportional to $e^{2S(m)}$ dominates the denominator and suppresses the wavefunction at $z=0$. Resonant enhancement occurs when its coefficient vanishes at $m=m_{\rm res}$:

\be \sin^2\Sigma(m_{\rm res})=0\,\, \implies \Sigma(m_{\rm res})=n\pi. \ee
Expanding near $m=m_{\rm res}$ gives the Breit--Wigner form

\begin{align}
&|\Psi_m(0)|^2=\frac{g_0^2}{\pi}
\frac{\Gamma_{\rm WKB}/2}{(m-m_{\rm res})^2+(\Gamma_{\rm WKB}/2)^2}, \\
&g_0^2=\frac{m_{\rm res}}{p(0)\Theta'(m_{\rm res})},
\qquad
\Gamma_{\rm WKB}=\frac{e^{-2S(m_{\rm res})}}{2\Theta'(m_{\rm res})}.
\end{align}

For the rectangular model shown schematically in Figure~\ref{fig:wkb-regions}, the potential takes the constant values $V_c$ in the central well and $V_b$ in the right barrier. The phase derivative and barrier action reduce to
\begin{equation}
    \Theta'(m_{\rm res})=
    \frac{m_{\rm res}z_3}{\sqrt{m_{\rm res}^2-V_c}},
    \qquad
    S(m_{\rm res})=(z_4-z_3)\sqrt{V_b-m_{\rm res}^2}.
\end{equation}
Substitution into the expression for $\Gamma_{\rm WKB}$ yields
\begin{equation}
    \Gamma_{\rm WKB}=
    \frac{1}{2z_3}\sqrt{1-\frac{V_c}{m_{\rm res}^2}}
    \exp\!\left[-2(z_4-z_3)\sqrt{V_b-m_{\rm res}^2}\right].
    \label{eq:wkb-rectangular-width}
\end{equation}

Although this model is only a qualitative approximation to the thick-brane effective potential, it captures the essential tunneling physics in a closed form analytic expression. As $\bar v$ decreases, the numerical potential develops less penetrable barriers, corresponding in the model to an increase of the barrier action $S$. Equation~\eqref{eq:wkb-rectangular-width} then gives an exponentially decreasing width, $\Gamma_{\rm WKB}\propto e^{-2S}$, so that the quasi-bound resonance continuously approaches a bound state in the flat-space limit. This provides the analytic counterpart of the narrowing resonance peaks found numerically in Section~\ref{sec:d}.

\section{Implications for the four-dimensional universe}\label{sec:implications}
The long-lived resonance motivates a discussion of its possible role in four-dimensional cosmology. We first summarize the time evolution of the shape-mode energy as it escapes from the thick brane into the bulk. We then use the directly fitted localized-energy leakage rate to examine whether couplings to brane-localized matter can transfer this energy to the four-dimensional sector and support high-scale reheating.

\paragraph{Evolution of energy density}

As in Section~\ref{sec:f}, $t$ and $z$ below denote dimensionless numerical coordinates. For the fluctuation $f(t,z)$, homogeneous along the three spatial brane directions, we define the energy density and the energy localized near the brane as
\begin{align}
  \rho(t,z)&=\frac{1}{2}\left[(\partial_t f)^2+(\partial_z f)^2+\bar V_s(z)f^2\right],
  \label{eq:implications-energy}\\
  E_c(t)&=\int_{-z_c}^{z_c}\rho(t,z)\,dz.
\end{align}
Here $z_c$ is chosen outside the outer classical turning point but well inside the taper and sponge regions. Thus $E_c(t)$ measures the energy remaining in the quasi-localized component rather than the total energy still present anywhere in the computational domain. We follow $E_c(t)/E_c(0)$ to quantify leakage from the near-brane region, and vary $z_c$ to check that the inferred rate is insensitive to the precise integration radius.

\begin{figure}[htbp]
    \centering
    \captionsetup[subfigure]{font=small}
    \begin{subfigure}{0.32\linewidth}
        \centering
        \includegraphics[width=\linewidth,height=3.4cm,keepaspectratio]{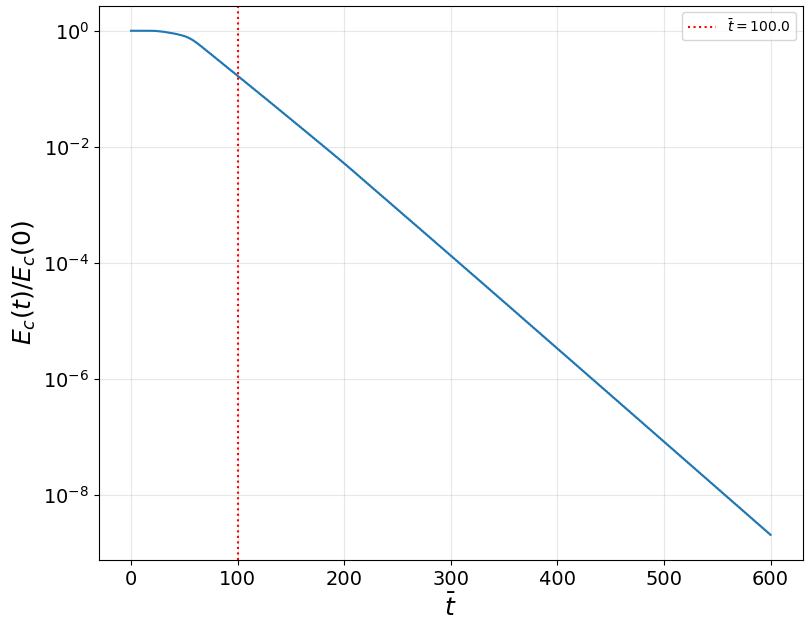}
        \caption{$\bar v=1.21$}\label{fig:implications-energy-121}
    \end{subfigure}%
    \hfill
    \begin{subfigure}{0.32\linewidth}
        \centering
        \includegraphics[width=\linewidth,height=3.4cm,keepaspectratio]{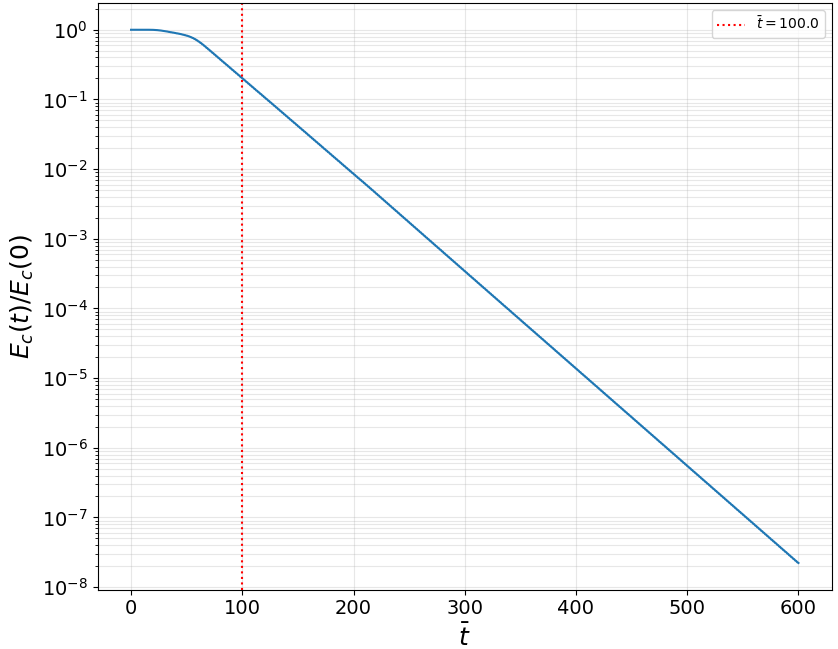}
        \caption{$\bar v=1.19$}\label{fig:implications-energy-119}
    \end{subfigure}%
    \hfill
    \begin{subfigure}{0.32\linewidth}
        \centering
        \includegraphics[width=\linewidth,height=3.4cm,keepaspectratio]{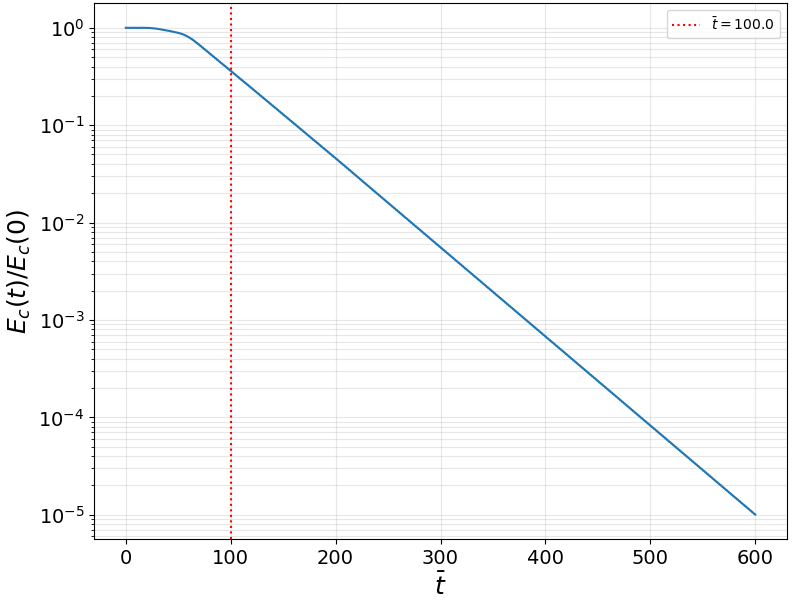}
        \caption{$\bar v=1.1$}\label{fig:implications-energy-11}
    \end{subfigure}
    \par\medskip
    \begin{subfigure}{0.32\linewidth}
        \centering
        \includegraphics[width=\linewidth,height=3.4cm,keepaspectratio]{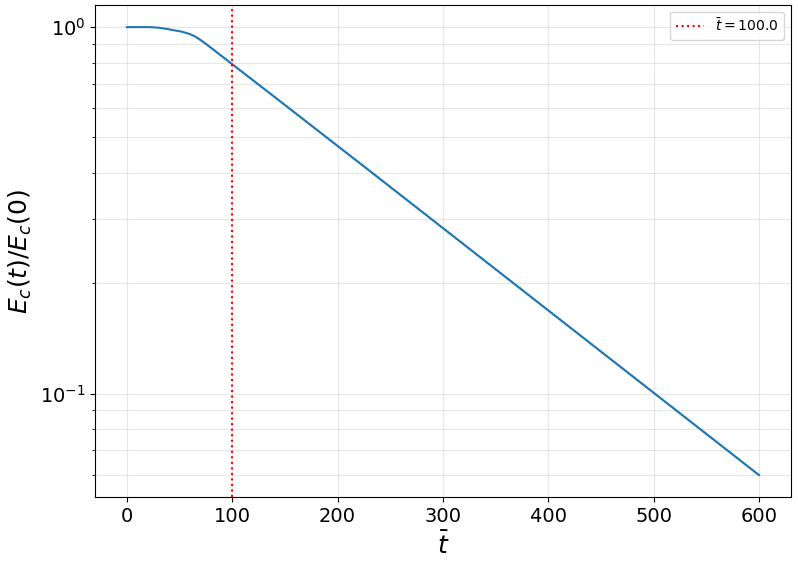}
        \caption{$\bar v=0.9$}\label{fig:implications-energy-09}
    \end{subfigure}%
    \hspace{0.02\linewidth}
    \begin{subfigure}{0.32\linewidth}
        \centering
        \includegraphics[width=\linewidth,height=3.4cm,keepaspectratio]{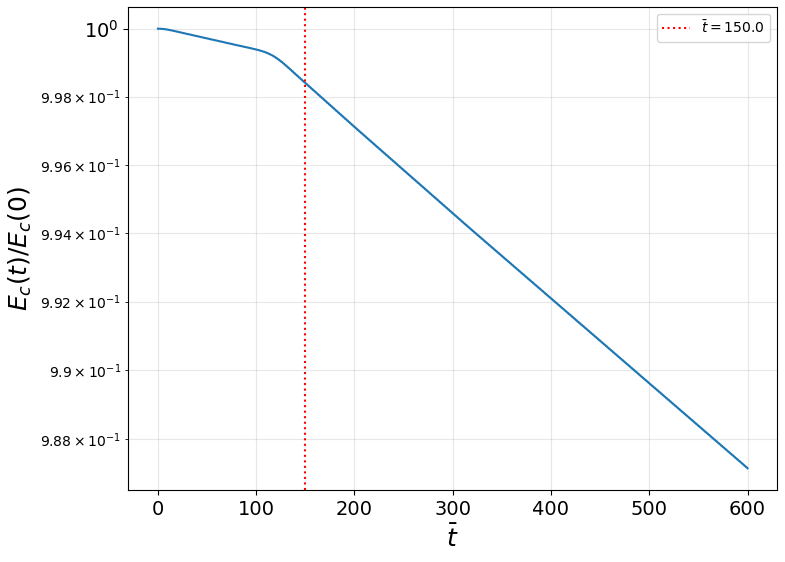}
        \caption{$\bar v=0.6$}\label{fig:implications-energy-06}
    \end{subfigure}
    \caption{Normalized energy $E_c(t)/E_c(0)$ for the five indicated values of $\bar v$. After the initial transient marked by the red dotted line, the late-time evolution exhibits clean exponential decay.}
    \label{fig:implications-energy}
\end{figure}

After the initial transient, the energy curves in Figure \ref{fig:implications-energy} are fitted using
\begin{equation}
  E_c(t)=A e^{-s(t-t_0)}+E_{\rm floor},\qquad
  \bar\tau_E=\frac{1}{s},\qquad
  \bar t_{1/2}=\frac{\ln2}{s}.
  \label{eq:implications-decay-times}
\end{equation}
Here $A$ is a free normalization, $E_{\rm floor}$ allows for a small late-time numerical or nonresonant remainder, and $s$ is the dimensionless energy-decay rate. The fit window is selected from a plateau of the instantaneous logarithmic slope $-d\ln E_c/dt$ and is varied as part of the convergence analysis. This avoids fitting either the initial readjustment or the very late numerical tail.

\begin{table}[htbp]
    \centering
    \caption{Energy-decay parameters of the shape mode: the fit coefficient $s$ and the dimensionless energy half-life.}
    \label{tab:lump-brane-summary}
    \begin{tabular}{|c|c|c|c|}
        \hline
        \textbf{$\bar v$} & \textbf{$\bar m_{\rm res}$} & \textbf{$s$} & \textbf{$\bar t_{1/2}$} \\
        \hline
        1.21 & 1.75166 & $3.51\times 10^{-2}$ & $1.97\times 10^{1}$ \\
        \hline
        1.19 & 1.75153 & $3.17\times 10^{-2}$ & $2.18\times 10^{1}$ \\
        \hline
        1.10 & 1.74895 & $2.08\times 10^{-2}$ & $3.32\times 10^{1}$ \\
        \hline
        0.9 & 1.74583 & $5.16\times 10^{-3}$ & $1.34\times 10^{2}$ \\
        \hline
        0.6 & 1.74039 & $1.65\times 10^{-4}$ & $4.19\times 10^{3}$ \\
        \hline
    \end{tabular}
\end{table}

The relative-probability width $\bar\Gamma_{\rm spec}$ provided a spectral estimate of the resonance timescale, but it was extracted from a finite-region localization diagnostic and need not coincide exactly with the decay rate measured in direct evolution. The two methods show the same strong suppression of leakage as $\bar v$ decreases, while their numerical rates differ by factors of order unity. We therefore use $\bar\Gamma_{\rm spec}^{-1}$ as an indicative localization timescale and regard the fitted coefficient $s$, obtained directly from the decay of $E_c(t)$, as the more precise measure of the physical bulk energy-loss rate wherever time-domain data are available.

\subsection{High-scale reheating}

Consider matter fields localized near the brane center $z=0$, with five-dimensional energy--momentum tensor $T_{MN}^{\rm matter}$. There are several ways to localize scalar and fermionic fields by appropriately coupling them to the brane-generating scalar field, see \cite{Liu:2017gcn} and references therein for more details. The interaction lagrangian density obtained by varying the matter action with respect to the metric perturbation $h_{MN}=G_{MN}-G_{MN}^{(0)}$ is
\begin{equation}
    \mathcal{L}_{\rm int}
    =-\frac{1}{2}\sqrt{-G^{(0)}}\,h^{MN}T_{MN}^{\rm matter}.
\end{equation}
In the longitudinal gauge, $\phi=-2\psi$, $h_{\mu\nu}=2e^{2A}\psi\eta_{\mu\nu}$, and $h_{zz}=-4e^{2A}\psi$. Using $\sqrt{-G^{(0)}}=e^{5A(z)}$, the interaction becomes
\begin{equation}
    \mathcal{L}_{\rm int}
    =-e^{3A(z)}\psi(x^\mu,z)
    \left[T^{\nu}{}_{\nu,{\rm matter}}(x^\mu,z)
    -2T^{zz}_{\rm matter}(x^\mu,z)\right].
\end{equation}

Because the extra dimension is noncompact and $V_s(z)\to0$ as $|z|\to\infty$, the spectrum is continuous. We therefore use the following continuum KK decomposition for the field $F(x^{\mu},z)$:
\begin{equation}
    F(x^\mu,z)=\int_0^\infty dm\,\chi_m(x^\mu)f_m(z),
    \label{eq:F_expansion}
\end{equation}
where
\begin{equation}
    -f_m''(z)+V_s(z)f_m(z)=m^2f_m(z),
    \label{eq:Schrodinger}
\end{equation}
and the continuum modes satisfy
\begin{equation}
    \int_{-\infty}^{\infty}dz\,f_m(z)f_{m'}(z)
    =k\,\delta(m-m').
    \label{eq:orthonormality}
\end{equation}
This convention is related to the delta-normalized WKB wavefunction by
$f_m(z)=\sqrt{k}\,\Psi_m(z)$ where $k$ is the inverse kink width.

\noindent For matter localized near the brane, we take
\begin{align}
    T^{\nu}{}_{\nu,{\rm matter}}^{(5{\rm D})}(x^\mu,z)
    &=T^{\nu}{}_{\nu,{\rm matter}}^{(4{\rm D})}(x^\mu)\Lambda(z),\\
    T^{zz}_{\rm matter}&\simeq0,
\end{align}
where $\Lambda(z)$ is peaked near $z=0$ and normalized by
\begin{equation}
    \int_{-\infty}^{\infty}dz\,e^{3A(z)}\Lambda(z)=1.
\end{equation}
For generic continuum masses, the potential barriers suppress the wavefunction near the brane. In the vicinity of $m=m_{\rm res}$, the wavefunction is instead resonantly enhanced within the central well. To describe the resonance by a canonically normalized four-dimensional field, we form a normalized wavepacket over a mass interval of width $\Gamma_{\rm bulk}^{({\rm spec})}$ centered on $m_{\rm res}$. Its extra-dimensional profile is
\begin{equation}
    u_{\rm res}(z)\equiv
    \frac{1}{\sqrt{k\Gamma_{\rm bulk}^{({\rm spec})}}}
    \int_{\rm res}dm\,f_m(z)
    \simeq\sqrt{\frac{\Gamma_{\rm bulk}^{({\rm spec})}}{k}}
    f_{m_{\rm res}}(z),
    \qquad
    \int_{-\infty}^{\infty}dz\,u_{\rm res}^2(z)=1.
\end{equation}
In the narrow-width approximation the field $F$ can therefore be written as
\begin{equation}
    F(x^\mu,z)\simeq Q(x^\mu)u_{\rm res}(z),
\end{equation}
where $Q$ is the effective canonically normalized four-dimensional resonance field.

The effective interaction is therefore
\begin{equation}
    S_{\rm int}^{(4{\rm D})}
    \simeq-\int d^4x\,g_QQ(x^\mu)
    T^{\nu}{}_{\nu,{\rm matter}}^{(4{\rm D})}(x^\mu),
\end{equation}
where
\begin{equation}
    g_Q=\frac{1}{M_*^4}\int_{-\infty}^{\infty}dz\,
    e^{\frac{3}{2}A(z)}\varphi_0'(z)
    u_{\rm res}(z)\Lambda(z)
    \simeq\frac{\varphi_0'(0)}{M_*^4}u_{\rm res}(0).
    \label{eq:reheating-effective-coupling}
\end{equation}
The factor $M_*^{-4}$ follows from the field redefinition in equation \eqref{eq:decomp}. For a sharply localized matter profile, the overlap is controlled by the normalized resonance wavepacket at the brane center, as shown in equation \eqref{eq:reheating-effective-coupling}. Because the interaction is proportional to ${T^\mu}_\mu=-\rho+3P$, the shape mode couples directly to nonrelativistic matter, whereas its linear coupling to an ideal conformal radiation fluid vanishes. It may nevertheless produce massive brane fields, which can subsequently decay or thermalize into radiation.

\paragraph{Quantitative estimates and reheating temperature:}
Two physically distinct decay rates enter this mechanism. We denote the physical escape rate into the noncompact bulk by $\Gamma_{\rm bulk}$; the superscripts $(E)$, $({\rm spec})$, and ${\rm WKB}$ distinguish the time-domain, spectral, and WKB estimates used below. Reheating is instead controlled by the decay rate into matter localized on the brane, denoted by $\Gamma_{\rm br}$. The total rate and the on-brane branching fraction are
\begin{equation}
    \Gamma_{\rm tot}=\Gamma_{\rm br}+\Gamma_{\rm bulk},
    \qquad
    B_{\rm br}=\frac{\Gamma_{\rm br}}{\Gamma_{\rm tot}}.
    \label{eq:reheating-branching-ratio}
\end{equation}
Efficient reheating requires
\begin{equation}
    \Gamma_{\rm br}\gg\Gamma_{\rm bulk},
    \label{eq:reheating-dominance-condition}
\end{equation}
so that the shape mode produces on-brane matter before most of its energy leaks into the bulk. The fit in equation \eqref{eq:implications-decay-times} gives the dimensionless energy-loss coefficient $s$. Restoring the kink scale, the corresponding physical bulk energy-loss rate is
\begin{equation}
    \Gamma_{\rm bulk}^{(E)}=s\,k.
    \label{eq:physical-bulk-width}
\end{equation}
The kink scale is actually related to the asymptotic AdS curvature of the five-dimensional spacetime. From equation \eqref{eq:Asol},
\begin{equation}
    A(y)\simeq-\kappa_{\rm AdS}|y|,
    \qquad
    \kappa_{\rm AdS}=\frac{\bar v^2}{18}k,
    \label{eq:ads-kink-scale-relation}
\end{equation}
where $\bar v=v/M_*^{3/2}$. However, fixing the asymptotic curvature does not independently fix the kink scale $k$ completely. On the other hand, the effective description requires the characteristic wall and resonance scales to remain below the five-dimensional cutoff. In particular,
\begin{equation}
    k\lesssim M_*,
    \qquad
    m_{\rm res}=\bar m_{\rm res}k\lesssim M_*.
    \label{eq:reheating-eft-conditions}
\end{equation}
The second condition is the stronger one for the resonance studied here, since $\bar m_{\rm res}\simeq\sqrt{3}$.

If the resonance dominates the energy density when it decays, its decay into brane matter dominates over bulk leakage, and the decay products thermalize rapidly, reheating occurs when $H\simeq\Gamma_{\rm br}$. With the Einstein--Hilbert normalization used in equation~\eqref{eq:Mpl}, this gives \cite{Kolb:1990vq,Allahverdi:2010xz}
\begin{equation}
    T_{\rm rh}\simeq\sqrt{\Gamma_{\rm br}M_{\rm Pl}},
    \label{eq:brane-reheating-temperature}
\end{equation}
where $M_{\rm Pl}\simeq1.22\times10^{18}\,\mathrm{GeV}$ in the convention of equation~\eqref{eq:Mpl}. Since $\Gamma_{\rm br}>\Gamma_{\rm bulk}$ is necessary for energy transfer to the brane to outpace bulk leakage, the bulk rate gives the conditional inequality
\begin{equation}
    T_{\rm rh}>
    \sqrt{\Gamma_{\rm bulk}M_{\rm Pl}}.
    \label{eq:reheating-lower-bound}
\end{equation}
This is not by itself a prediction of the reheating temperature: the actual value depends on the brane decay width, the initial abundance of the resonance, and the subsequent thermalization history. It does show that smaller values of $\bar v$, which strongly suppress bulk leakage, reduce the minimum brane decay rate needed for efficient energy transfer. The resulting scale can be compatible with the temperature required for standard thermal leptogenesis, $T_{\rm rh}\gtrsim10^9\,\mathrm{GeV}$, once a complete cosmological history is specified \cite{Davidson:2002qv,Buchmuller:2005eh}.

The explicit WKB calculation gives the dimensionless bulk width
\begin{equation}
    \bar\Gamma_{\rm bulk}^{\rm WKB}
    =\frac{e^{-2S(\bar m_{\rm res})}}
    {2\Theta'(\bar m_{\rm res})},
    \qquad
    \Gamma_{\rm bulk}^{\rm WKB}=k\bar\Gamma_{\rm bulk}^{\rm WKB},
    \label{eq:reheating-wkb-bulk-width}
\end{equation}
where $S$ and $\Theta$ are dimensionless functions of the dimensionless mass $\bar m=m/k$. At the resonance peak,
\begin{equation}
    \left|f_{m_{\rm res}}(0)\right|^2
    =4k\frac{\bar m_{\rm res}}{\pi\bar p(0)}
    e^{2S(\bar m_{\rm res})}.
\end{equation}
Here $\bar p(0)\equiv p(0)/k$ is the dimensionless local WKB wave number at the brane center.
Combining the peak amplitude with the normalized wavepacket factor and identifying the narrow spectral width with the WKB width gives
\begin{equation}
    \left|u_{\rm res}(0)\right|^2
    =\bar\Gamma_{\rm bulk}^{\rm WKB}
    \left|f_{m_{\rm res}}(0)\right|^2
    =\frac{2k\bar m_{\rm res}}
    {\pi\bar p(0)\Theta'(\bar m_{\rm res})},
\end{equation}
and hence
\begin{equation}
    g_Q^2=\frac{\bigl(\varphi_0'(0)\bigr)^2}{M_*^8}
    \frac{2k\bar m_{\rm res}}
    {\pi\bar p(0)\Theta'(\bar m_{\rm res})}.
    \label{eq:reheating-coupling-peak}
\end{equation}

\paragraph{Decay into brane scalars:}
For a canonically normalized, minimally coupled real brane scalar $H$, we use the canonical stress tensor
\begin{equation}
    T_{\mu\nu}^{(H)}=\partial_\mu H\partial_\nu H
    -\frac{1}{2}\eta_{\mu\nu}(\partial H)^2,
    \qquad
    {T^{(H)\mu}}_{\mu}=-(\partial H)^2.
\end{equation}
The trace interaction therefore has the derivative form
\begin{equation}
    \mathcal{L}_{\rm int}=g_QQ\,\partial_\mu H\partial^\mu H.
\end{equation}
The corresponding two-body decay width is
\begin{equation}
    \Gamma_{\rm br}^{(H)}
    =\frac{g_Q^2m_{\rm res}^3}{32\pi}.
    \label{eq:reheating-scalar-width}
\end{equation}
Using equation~\eqref{eq:reheating-coupling-peak}, this becomes
\begin{equation}
    \Gamma_{\rm br}^{(H)}
    =\frac{\bigl(\varphi_0'(0)\bigr)^2m_{\rm res}^3k}
    {16\pi^2M_*^8}
    \frac{\bar m_{\rm res}}
    {\bar p(0)\Theta'(\bar m_{\rm res})}.
\end{equation}
The ratio to the physical WKB leakage rate is therefore
\begin{equation}
    \frac{\Gamma_{\rm br}^{(H)}}{\Gamma_{\rm bulk}^{\rm WKB}}
    =\frac{\bigl(\varphi_0'(0)\bigr)^2m_{\rm res}^3}
    {8\pi^2M_*^8}
    \frac{\bar m_{\rm res}}{\bar p(0)}
    e^{2S(\bar m_{\rm res})}.
    \label{eq:reheating-width-ratio}
\end{equation}
Using $\varphi_0'(0)=vk$, $v=\bar v M_*^{3/2}$, and $m_{\rm res}=\bar m_{\rm res}k$, equation~\eqref{eq:reheating-width-ratio} can be written as
\begin{equation}
    \frac{\Gamma_{\rm br}^{(H)}}{\Gamma_{\rm bulk}^{\rm WKB}}
    =C(\bar v)\left(\frac{k}{M_*}\right)^5,
    \qquad
    C(\bar v)\equiv
    \frac{\bar v^2\bar m_{\rm res}^4}
    {8\pi^2\bar p(0)}e^{2S(\bar m_{\rm res})}.
    \label{eq:reheating-width-ratio-scaled}
\end{equation}
The inverse tunneling factor enhances the brane branching fraction, but equation~\eqref{eq:reheating-width-ratio-scaled} also displays the important suppression by $(k/M_*)^5$. Brane decay therefore dominates only within a parameter window in which the tunneling enhancement compensates for the cutoff suppression. Requiring a hierarchy $\Gamma_{\rm br}^{(H)}/\Gamma_{\rm bulk}^{\rm WKB}\geq R_{\min}$ while keeping the resonance below the cutoff gives
\begin{equation}
    \left(\frac{R_{\min}}{C(\bar v)}\right)^{1/5}
    \lesssim\frac{k}{M_*}
    \lesssim\frac{1}{\bar m_{\rm res}}.
    \label{eq:reheating-viable-window}
\end{equation}

For the numerical estimates, the turning points are obtained directly from the exact dimensionless potential by solving $\bar V_s(\bar z_i)=\bar m_{\rm res}^2$. We then evaluate
\begin{equation}
    \bar p(0)=\sqrt{\bar m_{\rm res}^2-\bar V_s(0)},
    \qquad
    S=\int_{\bar z_3}^{\bar z_4}
    \sqrt{\bar V_s(\bar z)-\bar m_{\rm res}^2}\,d\bar z,
    \label{eq:reheating-numerical-wkb-inputs}
\end{equation}
using the resonance masses in Table~\ref{tab:relProb-summary}. Thus the rectangular barrier introduced in Section~\ref{sec:add1} is used only to illustrate the WKB scaling; the numerical values in Table~\ref{tab:reheating-window} are calculated with the full exact potential.

\begin{table}[htbp]
    \centering
    \small
    \setlength{\tabcolsep}{5pt}
    \caption{WKB quantities entering the brane-to-bulk decay ratio and the EFT-consistent interval in $k/M_*$ for which $\Gamma_{\rm br}^{(H)}/\Gamma_{\rm bulk}^{\rm WKB}\geq10$. The upper endpoint follows from $m_{\rm res}<M_*$.}
    \label{tab:reheating-window}
    \begin{tabular}{|c|c|c|c|c|}
        \hline
        $\bar v$ & $\bar p(0)$ & $S$ & $C(\bar v)$ & Allowed $k/M_*$ interval \\
        \hline
        $0.6$ & $0.9612$ & $4.408$ & $2.93\times10^2$  & $0.509\lesssim k/M_*\lesssim0.575$ \\
        $0.5$ & $0.9738$ & $6.574$ & $1.52\times10^4$ & $0.231\lesssim k/M_*\lesssim0.575$ \\
        $0.4$ & $0.9835$ & $10.566$ & $2.82\times10^7$ & $0.051\lesssim k/M_*\lesssim0.576$ \\
        $0.3$ & $0.9909$ & $19.194$ & $4.89\times10^{14}$ & $0.0018\lesssim k/M_*\lesssim0.577$ \\
        \hline
    \end{tabular}
\end{table}

\noindent The table shows that the viable interval is narrow and close to the cutoff at $\bar v=0.6$, but expands rapidly as the gravitational coupling is reduced. The values $\bar v\gtrsim0.9$ do not admit a brane-to-bulk hierarchy of ten while simultaneously satisfying $m_{\rm res}<M_*$. An illustrative EFT-compatible benchmark is therefore $\bar v=0.4$, for which $C(0.4)\simeq2.8\times10^7$. Choosing
\begin{equation}
    \bar v=0.4,
    \qquad
    \frac{k}{M_*}=0.1
    \label{eq:reheating-controlled-benchmark}
\end{equation}
yields
\begin{equation}
    \frac{\Gamma_{\rm br}^{(H)}}{\Gamma_{\rm bulk}^{\rm WKB}}
    \simeq2.8\times10^2,
    \qquad
    \frac{m_{\rm res}}{M_*}\simeq0.17,
    \qquad
    \frac{\kappa_{\rm AdS}}{M_*}\simeq8.9\times10^{-4}.
    \label{eq:reheating-controlled-benchmark-results}
\end{equation}
Thus the decay into localized matter can dominate by more than two orders of magnitude while both the resonance mass and the background curvature remain below the five-dimensional cutoff. More generally, for $\bar v=0.4$ the requirement of a ten-to-one hierarchy, $R_{\min}=10$, gives approximately
\begin{equation}
    0.051\lesssim\frac{k}{M_*}\lesssim0.576.
\end{equation}
The benchmark lies within this interval and remains substantially below the resonance-mass cutoff at its upper endpoint. Most importantly, the allowed interval widens substantially at weak gravitational coupling, corresponding to smaller $\bar v$, and is no longer restricted to a narrow region near the cutoff. Thus, for $\bar v\lesssim0.5$, there is a broad EFT-consistent parameter range in which decay into localized matter dominates bulk leakage. Provided that the resonance has a cosmologically significant abundance and its decay products thermalize efficiently, this range can facilitate high-scale reheating. Its ultraviolet radiative stability, however, depends on the microscopic completion of the scalar and matter-localization sectors.

\section{Summary and conclusions}\label{sec:conclusions}

We have investigated the fate of the internal shape mode of the flat-space $\varphi^4$ kink when the kink is promoted to a self-gravitating thick brane in five-dimensional Einstein-scalar theory. At finite gravitational coupling, the coupled scalar-metric fluctuation's effective potential approaches zero asymptotically and therefore does not support a normalizable massive bound state. Nevertheless, the relative-probability spectrum contains a Breit-Wigner peak hinting the existence of a single quasi-localized (resonant) mode. The mass of the resonant mode approaches $\bar m^2=3$ and its reconstructed scalar-field profile approaches the flat-space shape-mode wavefunction as $\bar v$ decreases, thus numerically showing that this resonant scalar mode is the gravitationally dressed remnant of the flat-space bound kink shape mode.
\noindent The time-domain evolution provides an independent dynamical characterization of this state. Initial data constructed from the resonant profile exhibit oscillations at the frequency predicted by the time-independent Schr\"{o}dinger problem and lose energy from within the thick-brane region through radiation into the bulk. The relative-probability width and the localized-energy leakage rate show the same strong suppression as $\bar v$ decreases, although they differ quantitatively by factors of order unity. We therefore treat them as distinct operational estimators rather than identifying either one directly with the complex pole width. The four-turning-point WKB analysis reproduces the resonance condition and shows that barrier penetration is controlled by the factor $e^{-2S}$. The increasing barrier action in the weak-gravity regime consequently explains the rapid narrowing and slower leakage of the resonance, together with the recovery of a stable bound shape mode in the flat-space limit.

Because the resonance appears as a brane-localized metric perturbation, it couples to the trace of the stress tensor of matter localized within the brane thickness. For the canonical scalar channel considered here, the brane-to-bulk width ratio contains both the inverse tunneling enhancement and the cutoff suppression $(k/M_*)^5$. Decay into localized matter is therefore not automatically dominant. Nevertheless, we identified a finite EFT-consistent parameter window in which the brane decay width exceeds the bulk-leakage width while the wall scale, resonance mass, and AdS curvature remain below the five-dimensional cutoff. This establishes the viability, rather than a complete prediction, of post-inflationary reheating sourced by the resonant scalar-metric fluctuation. The resulting reheating temperature remains conditional on the resonance abundance, its absolute brane decay width, and efficient thermalization of the decay products.

A more complete cosmological treatment should incorporate an expanding brane background, a production mechanism and initial abundance for the resonance, a specified localization mechanism for the full matter sector, and the subsequent nonequilibrium thermalization of the decay products. Another interesting direction is to investigate whether the scalar field in this model can be interpreted as an Einstein-frame scalaron arising from a consistent $f(R)$ theory.

\section*{Acknowledgments}
GC would like to acknowledge very useful discussions with Prof. Sourov Roy, Prof. Arnab Das and Prof. Sumanta Chakraborty during the preparation of this manuscript. GC is supported by DST-IACS Masters Fellowship.

\appendix
\section{Numerical convergence tests}
\label{app:numerical-convergence}

This appendix presents the numerical validation of the spectral widths and time-domain decay rates for two representative couplings, $\bar v=1.10$ and $\bar v=0.60$. The former has a comparatively broad, short-lived resonance, whereas the latter has a narrow resonance that requires a substantially longer evolution. These cases therefore probe complementary numerical regimes. Unless stated otherwise, each parameter is varied independently while the remaining parameters are held at their baseline values.

\subsection{Convergence of the spectral width}

The relative-probability peak is first located on a coarse grid and then recomputed on a fine mass grid centered on the peak. We extract the full width at half prominence, $\bar\Gamma_{\rm HP}$, and independently fit
\begin{equation}
 P_{\rm fit}(\bar m)=B_0+B_1(\bar m-\bar m_{\rm res})
 +\frac{C}{(\bar m-\bar m_{\rm res})^2+(\bar\Gamma_{\rm BW}/2)^2},
 \label{eq:appendix-bw-fit}
\end{equation}
where the linear term accounts for the slowly varying nonresonant background. The fine-grid spacing and fitting interval are reduced until the fitted mass and width cease to change appreciably. The final results are given in Table~\ref{tab:appendix-spectral-convergence}.

\begin{table}[htbp]
  \centering
  \caption{Static resonance extraction on the final fine mass grids. The last column gives the fractional difference between the half-prominence and Breit--Wigner widths.}
  \label{tab:appendix-spectral-convergence}
  \small
  \begin{tabular}{|c|c|c|c|c|c|}
    \hline
    $\bar v$ & $\Delta\bar m$ & $\bar m_{\rm res}$ & $\bar\Gamma_{\rm HP}$ & $\bar\Gamma_{\rm BW}$ & $|\Gamma_{\rm HP}/\Gamma_{\rm BW}-1|$ \\
    \hline
    1.10 & $4.59\times 10^{-5}$ & $1.74895$ & $2.91\times 10^{-2}$ & $2.96\times 10^{-2}$ & $0.016$ \\
    \hline
    0.60 & $2.5\times 10^{-6}$ & $1.74039$ & $3.76\times 10^{-4}$ & $3.78\times 10^{-4}$ & $0.005$ \\
    \hline
  \end{tabular}
\end{table}

For $\bar v=1.10$, the half-prominence and Breit--Wigner procedures give widths of $2.91\times10^{-2}$ and $2.96\times10^{-2}$, respectively, differing by $1.6\%$. For the much narrower $\bar v=0.60$ resonance, the corresponding widths are $3.76\times10^{-4}$ and $3.78\times10^{-4}$, differing by only $0.5\%$. The two independent estimators therefore resolve both the broad and narrow peaks consistently. In the main text we use the half-prominence width.

\subsection{Convergence of the time-domain decay rate}

\begin{figure}[htbp]
  \centering
  \begin{subfigure}{0.88\linewidth}
    \centering
    \IfFileExists{conv11.png}{%
      \includegraphics[width=\linewidth]{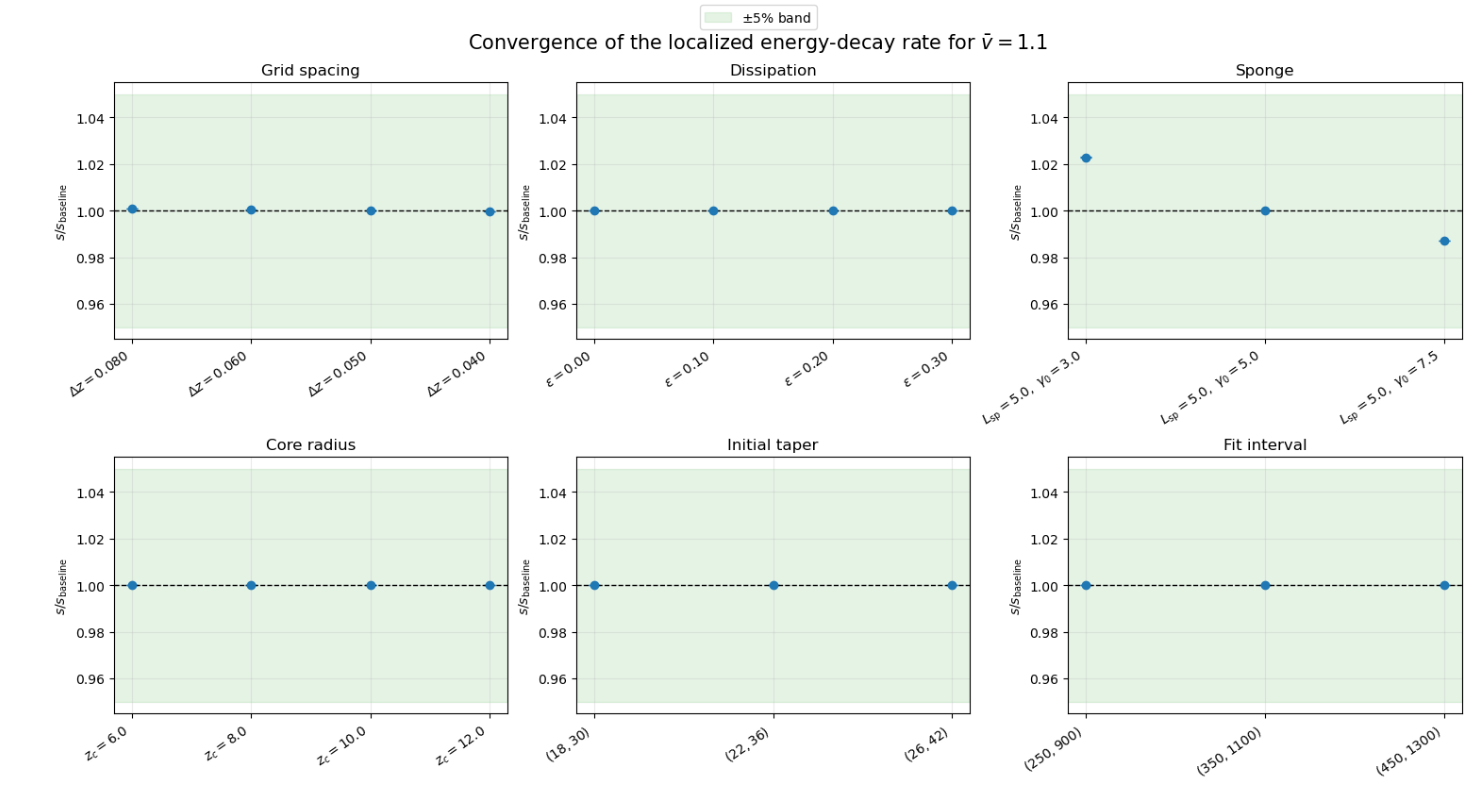}%
    }{%
      \fbox{\parbox[c][4.5cm][c]{0.9\linewidth}{\centering Insert the convergence plot for $\bar v=1.10$ here.}}%
    }
    \caption{$\bar v=1.10$}
  \end{subfigure}
  \par\medskip
  \begin{subfigure}{0.88\linewidth}
    \centering
    \IfFileExists{conv06.png}{%
      \includegraphics[width=\linewidth]{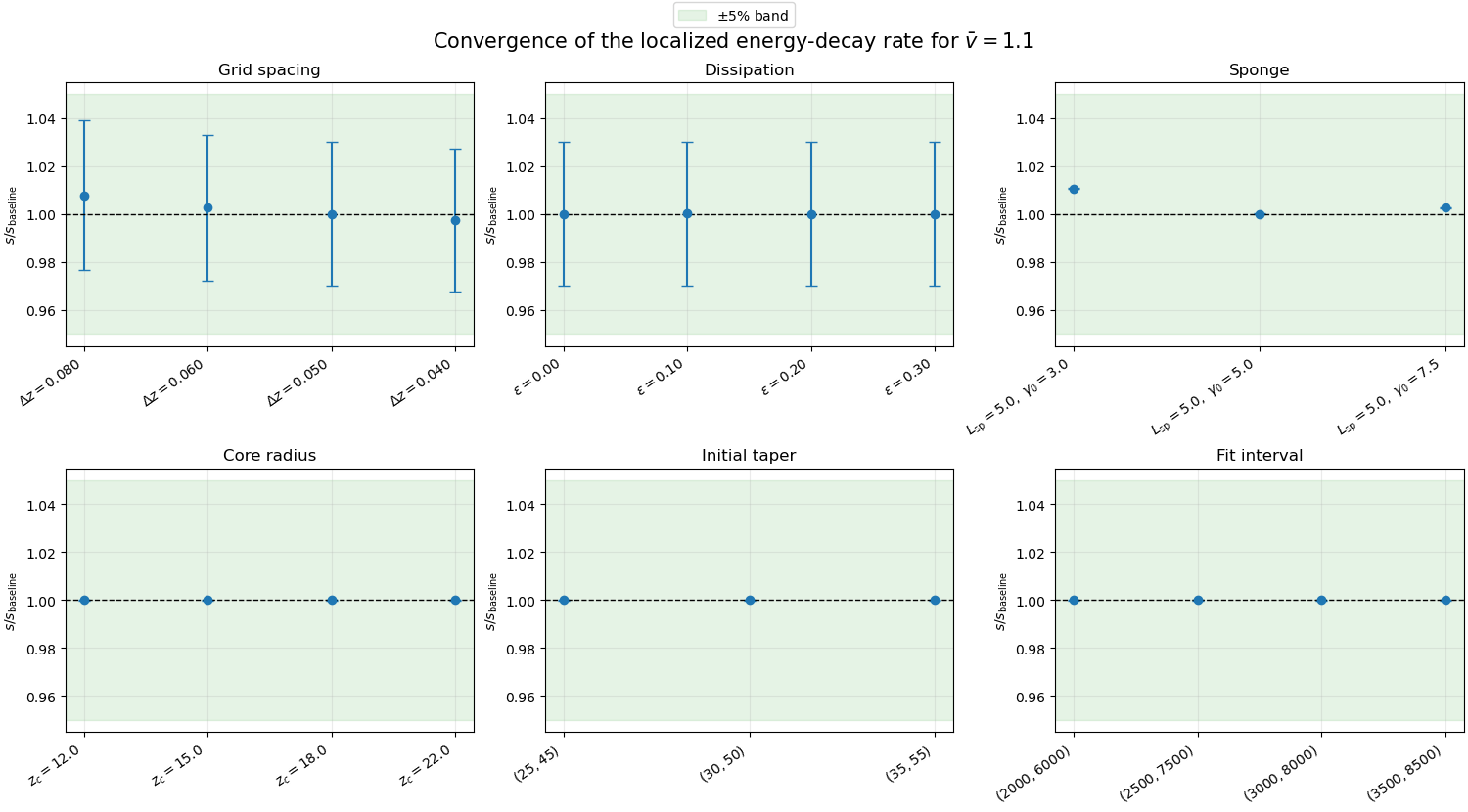}%
    }{%
      \fbox{\parbox[c][4.5cm][c]{0.9\linewidth}{\centering Insert the convergence plot for $\bar v=0.60$ here.}}%
    }
    \caption{$\bar v=0.60$}
  \end{subfigure}
  \caption{Normalized localized-energy decay rate $R_s=s/s_{\rm ref}$ under variations of the numerical parameters for (a) $\bar v=1.10$ and (b) $\bar v=0.60$. The dashed line denotes the baseline result and the shaded region is the adopted $\pm5\%$ tolerance band.}
  \label{fig:appendix-convergence}
\end{figure}

The decay rate is extracted from the localized energy $E_c(t)$ defined in equation~\eqref{eq:implications-energy}. We vary the spatial grid spacing $\Delta z$, the core radius $z_c$, the taper interval $(z_{\rm tap}^{\rm in},z_{\rm tap}^{\rm out})$, the fit window, the sponge parameters, and the Kreiss--Oliger coefficient. For each variation we quote the ratio
\begin{equation}
 R_s=\frac{s}{s_{\rm ref}},
 \label{eq:appendix-rate-ratio}
\end{equation}
where $s_{\rm ref}$ is obtained from the baseline run at the same value of $\bar v$. The core radius is always chosen beyond the outer turning point, while the taper and sponge begin farther from the brane. The shaded regions in Figure~\ref{fig:appendix-convergence} indicate a conservative $\pm5\%$ tolerance around the baseline. Error bars, where visible, are the uncertainties returned by the exponential fit.

Tables~\ref{tab:appendix-time-convergence-v11} and~\ref{tab:appendix-time-convergence-v06} summarize the tested parameter values and the maximum fractional deviation
\begin{equation}
 \delta_s^{\max}=\max_i\left|\frac{s_i}{s_{\rm ref}}-1\right|.
\end{equation}

\begin{table}[htbp]
  \centering
  \caption{Summary of the time-domain convergence tests for $\bar v=1.10$. The final column gives the maximum fractional deviation from the baseline decay rate.}
  \label{tab:appendix-time-convergence-v11}
  \small
  \begin{tabular}{|c|c|c|}
    \hline
    Test & Values varied & $\delta_s^{\max}$ \\
    \hline
    Grid spacing $\Delta z$ & $0.04,0.05,0.06,0.08$ & $6.3\times 10^{-5}$ \\
    \hline
    Core radius $z_c$ & $6,8,10,12$ & $1.1\times 10^{-3}$ \\
    \hline
    Taper interval & $(18,30),(22,36),(26,42)$ & $1.9\times 10^{-3}$ \\
    \hline
    Fit window & $(250,900),(350,1100),(450,1300)$ & $1.2\times 10^{-2}$ \\
    \hline
    Sponge $(\gamma_0,L_{\rm sp})$ & $(3,5),(5,5),(7.5,5)$ & $2.3\times 10^{-2}$ \\
    \hline
    KO coefficient $\epsilon$ & $0,0.1,0.2,0.3$ & $8.04\times 10^{-7}$ \\
    \hline
  \end{tabular}
\end{table}

\begin{table}[htbp]
  \centering
  \caption{Summary of the time-domain convergence tests for $\bar v=0.60$. The final column gives the maximum fractional deviation from the baseline decay rate.}
  \label{tab:appendix-time-convergence-v06}
  \small
  \begin{tabular}{|c|c|c|}
    \hline
    Test & Values varied & $\delta_s^{\max}$ \\
    \hline
    Grid spacing $\Delta z$ & $0.04,0.05,0.06,0.08$ & $7.8\times 10^{-3}$ \\
    \hline
    Core radius $z_c$ & $12,15,18,22$ & $5.6\times 10^{-7}$ \\
    \hline
    Taper interval & $(25,45),(30,50),(35,55)$ & $1.2\times 10^{-7}$ \\
    \hline
    Fit window & $(2000,6000),(2500,7500),(3000,8000),(3500,8500)$ & $1.7\times 10^{-5}$ \\
    \hline
    Sponge $(\gamma_0,L_{\rm sp})$ & $(3,5),(5,5),(7.5,5)$ & $1.06\times 10^{-2}$ \\
    \hline
    KO coefficient $\epsilon$ & $0,0.1,0.2,0.3$ & $2.1\times 10^{-4}$ \\
    \hline
  \end{tabular}
\end{table}

For $\bar v=1.10$, the largest variation is produced by the sponge parameters, but the maximum change is only $2.3\%$. Changing the fit interval produces a $1.2\%$ variation, while the grid spacing, core radius, taper interval, and numerical dissipation affect the rate at the sub-percent level. For $\bar v=0.60$, the sponge and grid tests give the largest deviations, $1.06\%$ and $0.78\%$, respectively. The remaining variations are at or below the $2.1\times10^{-4}$ level. All central values therefore remain well within the adopted $5\%$ tolerance.

The two cases require very different fit windows. For $\bar v=1.10$, the fit must begin after the initial adjustment but before the rapidly decaying resonant component reaches the numerical floor. For $\bar v=0.60$, the narrow resonance must instead be evolved to much later times before a sufficiently long decay interval is accumulated. The stability under changes of the core radius and taper interval shows that the fitted rate is not controlled by the arbitrary definition of the near-brane region or by truncation of the initial scattering profile. Likewise, the sponge and dissipation tests show that boundary absorption and high-frequency filtering do not determine the measured leakage rate. These results support the interpretation of the observed exponential decay as a physical loss of the quasi-localized mode into the bulk.

\bibliographystyle{biblio}
\bibliography{main}

\end{document}